\documentclass[aps,prd,twocolumn,nofootinbib,showpacs,superscriptaddress]{revtex4-1}
\usepackage{graphicx}
\usepackage[colorlinks=true,linkcolor=blue,citecolor=blue,urlcolor=blue]{hyperref}
\usepackage{slashed,color,amsmath,amssymb,mathrsfs}
\usepackage{multirow}
\usepackage{diagbox}
\usepackage{bbm}
\usepackage{longtable}
\usepackage{array}
\usepackage{footmisc}
\usepackage{extarrows}
\usepackage{autobreak}
\usepackage{xcolor}

\allowdisplaybreaks

\begin{document}

\title{Study of electron-positron annihilation into four pions within chiral effective field theory in the low energy region}

\author{Jia-Yu Zhou}
\affiliation{School for Theoretical Physics, School of Physics and Electronics, Hunan University, Changsha 410082, China}
\affiliation{Hunan Provincial Key Laboratory of High-Energy Scale Physics and Applications, Hunan University, Changsha 410082, People's Republic of China}

\author{Hao-Xiang Pan}
\affiliation{School for Theoretical Physics, School of Physics and Electronics, Hunan University, Changsha 410082, China}
\affiliation{Hunan Provincial Key Laboratory of High-Energy Scale Physics and Applications, Hunan University, Changsha 410082, People's Republic of China}

\author{Ling-Yun Dai}\email{dailingyun@hnu.edu.cn}
\affiliation{School for Theoretical Physics, School of Physics and Electronics, Hunan University, Changsha 410082, China}
\affiliation{Hunan Provincial Key Laboratory of High-Energy Scale Physics and Applications, Hunan University, Changsha 410082, People's Republic of China}

\begin{abstract}
In this paper, we employ chiral effective field theory to study the process of electron-positron annihilation into four pions in the low energy region within $E_{c.m.}\leq 0.6$ GeV. The prediction of the cross section is obtained through $SU(3)$ chiral perturbation theory up to the next-to-leading order, which is smaller than the experimental data in the energy region [0.6-0.65] GeV, though the data has only a few points and poor statistics. 
Then, the resonance chiral theory is applied to include the resonance contribution, with the lightest scalars and vectors written in the effective Lagrangians. A series of relevant decay widths and the masses of the vectors are studied to fix the unknown couplings. 
The resonance contribution should be one order larger than that of the chiral perturbation theory but still one to two orders smaller than the data. The significant discrepancy urged the new experimental measurements to give more guidance. We also compute the leading order hadronic vacuum polarization contribution from the four pion channels to the anomalous magnetic moment of the muon, $(g-2)_\mu$. In the energy range from threshold up to 0.6 GeV within resonance chiral theory, the contributions are $a_\mu=(0.680\pm0.062)\times10^{-16}$ and $a_\mu=(0.597\pm0.058)\times10^{-16}$ for the processes of $e^+e^-\to\pi^+\pi^+\pi^-\pi^-$, $\pi^0\pi^0\pi^+\pi^-$, respectively.
\end{abstract}

\maketitle

\section{Introduction}
Electron-positron annihilation processes provide a pristine experimental setting to investigate the properties and interactions of fundamental particles. In the low energy region, such processes with final states of the lightest hadrons have been precisely measured by many experiments. 
In contrast, they can hardly be described by theoretical models.
Quantum Chromodynamics (QCD), widely recognized as the basic theory of strong interaction, works well at high energies. However, the non-perturbative property of QCD makes it challenging to apply directly in the low-energy region. 
Chiral effective field theory (ChEFT) \cite{Weinberg:1978kz,Gasser:1983yg,Gasser:1984gg} can be considered as low energy effective field theory (EFT) of QCD with respect to the chiral symmetry of the light quarks. 
ChEFT is successful in describing interactions involving the lightest pseudoscalars. 
For example, chiral perturbation theory (ChPT) \cite{Weinberg:1978kz,Gasser:1983yg,Gasser:1984gg} works well on the scattering amplitudes between the lightest pseudoscalars, 
with $SU(2)$ ChPT focusing on the pions (up and down quarks) and $SU(3)$ ChPT extending to the kaons and eta (including the heavier strange quark).
Indeed, these amplitudes can be used to work through the final-state interactions (FSI) of the pseudoscalars for the electron-positron annihilation processes. They are the basics to construct the hadronic vacuum polarization form factors (HVP), which are essential to study the anomalous magnetic moment of the muon, $a_\mu=(g-2)_\mu/2$. 
The process of electron-positron annihilating into four pions makes significant contributions to HVP.

As is known, $(g-2)_\mu/2$ acts as one of the most precise indicators for new physics beyond the standard model (SM) \cite{Jegerlehner:2017gek}. 
The experimental measurements \cite{Muong-2:2006rrc,Muong-2:2021ojo,Muong-2:2023cdq} 
deviate 4-5 $\sigma$ from the Standard Model (SM) theoretical prediction. See Refs.~\cite{Aoyama:2020ynm} and references therein and thereafter. 
However, predictions from different theoretical models show significant discrepancies. Unlike the data-driven approach \cite{Keshavarzi:2018mgv,Colangelo:2019uex,Davier:2019can,Keshavarzi:2019abf,Qin:2020udp, Wang:2023njt, Fedotovich:2024dpk}, lattice QCD provides a much higher estimation of the HVP contribution \cite{Borsanyi:2020mff, Ce:2022kxy,ExtendedTwistedMass:2022jpw,FermilabLatticeHPQCD:2023jof,RBC:2023pvn, Erb:2024rob,Colangelo:2021moe}, resulting in a theoretical prediction that is much closer to the experimental value and, therefore, a much smaller discrepancy. Very recently, CMD-3 announced their latest measurement of the cross-section of $e^+e^- \to \pi\pi$~\cite{CMD-3:2023alj, CMD-3:2023rfe}, which is larger than the previous measurements and consistent with the results of the lattice QCD discussed above. 
Also, the HVP contribution from $\tau$ decay is larger than that of electron-positron annihilation but closer to that of lattice QCD \cite{Miranda:2018cpf,Hoferichter:2023sli,GomezDumm:2013sib}. Nevertheless, it should be noted that the HVP contribution derived from $\tau$ decay is subject to isospin-breaking contributions, which are currently a matter of considerable debate \cite{Aliberti:2025beg}.

In the present work, we restrict our analysis of the process of $ee\to4\pi$ from the threshold up to 0.6~GeV, and it is extended to 0.65~GeV to compare with the data.
ChPT is expected to work well in this energy region with the basic principles of quantum field theory (QFT). Notice that the upper limit, 0.6~GeV, is almost 0.178~GeV far away from the resonance $\rho(770)$ mass. The distance is larger than the width of the $\rho(770)$, and the effects from the resonance should be ignorable.
There have been previous works applying $SU(2)$ ChPT to study the process of electron-positron annihilation into four pions~\cite{Unterdorfer:2002zg,Ecker:2002cw}, up to next-to-leading order (NLO). It is found that the $SU(2)$ ChPT result can not match the data of $e^+e^-\to\pi^+\pi^+\pi^-\pi^-$ at 0.6-0.65~GeV. A natural thought is to apply the $SU(3)$ ChPT. For instance, the mass of kaon-antikaon is not far away from 0.65~GeV. 
However, as will be discussed in the following sections, the prediction of $SU(3)$ ChPT is similar to that of $SU(2)$ ChPT, two to three orders smaller than the data. 
To solve this problem, we use the resonance chiral theory (RChT) \cite{Ecker:1988te, Ecker:1989yg,Cirigliano:2006hb,Guo:2007ff,Portoles:2010yt,Dai:2013joa,Qin:2024ulb} to include the effects of the vectors and the lightest scalars.
The spirit of RChT is to include the resonances as new degrees of freedom, and the interaction Lagrangians between these resonances and the pseudoscalars are still constrained by the chiral symmetry as well as the discrete symmetries, e.g., the conservation of parity ($P$) and charge conjugation ($C$). Its power counting is compensated by large $N_C$ (the number of colors) expansions. 
To fix the unknown couplings in RChT, we also give a fit to the decay widths of $V\to e^+e^-$ and $V\to S\gamma$. Here, $V$ and $S$ are the lightest vectors and scalars, respectively. Then, with these fixed couplings, one can predict the cross section of $e^+e^-\to 4\pi$ and compare it with that of ChPT. 
As will be discussed, those resonances have non-ignorable contributions and enlarge the cross sections one to two orders. 

The structure of this paper is as follows: In Sec.~II, we give a brief introduction to the theory framework, ChPT and RChT.
In Sec.~III, the two ways to get the amplitudes are mentioned. In solution I, the scattering amplitudes of $e^+e^-\to 4\pi$ are calculated within the $SU(3)$ ChPT up to the next-to-leading order (NLO),
and in solution II, the scattering and decay amplitudes are calculated from RChT. 
In Sec.~IV, we give numerical results on the cross section of $e^+e^-\to 4\pi$, both from ChPT and RChT. The way to fix the unknown couplings of the effective Lagrangians is also discussed. 
Also, we estimate the HVP contributions of $e^+e^-\to 4\pi$ channel to the $(g-2)_\mu$. 
Finally, we summarize our results.

\section{Theory framework}
As was introduced in the previous section, there are two theoretical frameworks for ChEFT: one is from ChPT, and the other is from the RChT. 
Both of them are of the $SU(3)$ cases.
For the former, the low energy constants (LECs) are well fixed in the low energy region \cite{Bijnens:2014lea}, so we use these numbers directly. The result is named Solution I. 
For the latter, the unknown couplings are fixed by fitting to the masses of the lightest vector resonances and the decay widths of $V(\rho,\omega,\phi)\to e^+e^-$ and $V\to S(a_0,f_0)\gamma$. The result is called Solution II. 
Furthermore, we combine the contributions from ChPT and RChT to evaluate the individual effects from different components, e.g., the ChPT one-loop and RChT resonance parts. We incorporate the one-loop ChPT terms into the RChT framework, denoted as Solution III.

\subsection{Effective Lagrangian for ChPT} \label{sec:ChPT}
The SU(3) chiral symmetry group, $G = SU(3)_L\otimes SU(3)_R$, arises naturally in the chiral limit where the up, down, and strange quark masses are ignored. The transformations of the group act independently on the left- and right-handed quarks.
This symmetry is spontaneously broken to its subgroup, $SU(3)_V$, resulting in the emergence of eight Nambu-Goldstone bosons: the three pions $(\pi^+, \pi^-, \pi^0)$, the four kaons $(K^+, K^-, K^0,\bar{K}^0)$, and the eta ($\eta$).
The Goldstone fields, denoted as $\Phi$, characterize the components $u(\Phi)$ from the coset space $SU(3)_L \otimes SU(3)_R/SU(3)_V$. A specific representation of these components is as follows:
\begin{equation}
 u(\Phi)=exp\biggl\{\frac{i}{\sqrt{2}F}\Phi\biggr\},
\end{equation}
with
\begin{eqnarray}
\Phi=\left(\begin{array}{c c c}
\frac{\pi^0}{\sqrt{2}}+\frac{\eta_8}{\sqrt{6}}&\pi^+ & K^+ \\
\pi^-& -\frac{\pi^0}{\sqrt{2}}+\frac{\eta_8}{\sqrt{6}} & K^0 \\
 K^- &\bar{K}^0&-\frac{2\eta_8}{\sqrt{6}} \\
\end{array}
\right)\,,
\end{eqnarray}
and $F$ is the pion decay constant, $F\simeq F_\pi=92.2$ MeV \cite{ParticleDataGroup:2022pth}. 
The leading order (LO) Lagrangian for ChPT can be written as \cite{Gasser:1984gg}:
\begin{equation}
\mathcal{L}^{\rm ChPT}_{(2)}=\frac{F^2}{4}\bigl<u_\mu u^\mu+\chi_+\bigr>\ . \label{Eq:L;Op2}
\end{equation}
The NLO Lagrangians ($\mathcal{O}(p^4)$ with $p$ the external momenta) are given as \cite{Gasser:1984gg,Scherer:2012xha}:
\begin{eqnarray}
\mathcal{L}^{\rm ChPT}_{(4)}&=&L_{1}\big<u_{\mu}u^{\mu}\big>\big<u_{\nu}u^{\nu}\big>+L_{2}\big<u_{\mu}u_{\nu}\big>\big<u^{\mu}u^{\nu}\big> \nonumber\\
&\!+\!&\!L_{3}\!\big<\!u_{\mu}u^{\mu}u_{\nu}u^{\nu}\!\big>\!+\!L_{4}\big<u_{\mu}u^{\mu}\big>\!\big<\chi_{+}\big>\!+\!L_{5}\!\big<\!u_{\mu}u^{\mu}\chi_{+}\!\big> \nonumber\\
&\!+\!&\!L_{6}\big<\chi_{+}\big>^{2}+L_{7}\big<\chi_{-}\big>^{2}+\frac{L_{8}}{2}\big<\chi_{+}^{2}+\chi_{-}^{2}\big> \nonumber\\
&\!-\!&\!iL_{9}\big<f^{\mu\nu}_{+}u_{\mu}u_{\nu}\big>+\frac{L_{10}}{4}\big<f^{2}_{+}-f^{2}_{-}\big> \nonumber\\
&\!+\!&\!H_{1}\big<f^{R}_{\mu\nu}f^{\mu\nu}_{R}+f^{L}_{\mu\nu}f^{\mu\nu}_{L}\big>+H_{2}\big<\chi\chi^{\dagger}\big>,\label{Eq:L;Op4}
\end{eqnarray}
where one has 
\begin{eqnarray}
\chi&=&2B\hat{m}\mathbf{1}\\
\chi_\pm&=&u^\dagger\chi u^\dagger\pm u\chi^\dagger u \nonumber\\
 F^{\mu\nu}_L&=&\partial^\mu l^\nu-\partial^\nu l^\mu-i[l^\mu,l^\nu] \nonumber\\
 F^{\mu\nu}_R&=&\partial^\mu r^\nu-\partial^\nu r^\mu-i[r^\mu,r^\nu] \nonumber\\
 f^{\mu\nu}_\pm&=&uF^{\mu\nu}_Lu^\dagger\pm u^\dagger F^{\mu\nu}_Ru.
\end{eqnarray}
Here, B is the coupling relating to the quark condensate, $\hat{m}\mathbf{1}$ is the mass matrix, $\hat{m}\mathbf{1}=\rm{diag}(\hat{m_\pi}^2,\hat{m_\pi}^2,2\hat{m_K}^2-\hat{m_\pi}^2)$, and $L_{i}$, $H_{i}$ are the LECs. The external vector sources are responsible for producing electromagnetic form factors based on the given assignments, $r^\mu=l^\mu=-e \mathcal{Q} A_\mu$, where $\mathcal{Q}$ is the electric charge matrix, $\mathcal{Q}=\rm{diag}(2,-1,-1)/3$.

\subsection{Effective Lagrangians of RChT}
\label{sec:RChT}
As will be discussed in the next sections, the ChPT result (of $ee\to\pi\pi\pi\pi$ cross section) is much smaller than the data in the energy region of 0.6-0.65~GeV. The reason may be that the resonance contribution can not be ignored, especially the closest $\rho$ and $\sigma$. Therefore, we try another way, RChT, to study the dynamics of this process. The resonances are written explicitly in the effective Lagrangians, filled in the octets and singlets as
\begin{eqnarray}
 R = \sum_{i=1}^8 \frac{\lambda_i}{\sqrt{2}}R_i + \frac{R_0}{\sqrt{3}}\mathbbm{1}~,
\end{eqnarray}
where $R=V, S$ denotes the lightest vectors and scalars, respectively. 
The vector mesons are described by the anti-symmetric tensor field~\cite{Ruiz-Femenia:2003jdx,Dai:2013joa}, which can be filled in an explicit matrix form, 
\begin{eqnarray}
V_{\mu\nu}\!=\!
\left(\!
\begin{array}{c c c}
\frac{\rho^0}{\sqrt{2}}\!+\!\frac{\omega_8}{\sqrt{6}}\!+\!\frac{\omega_0}{\sqrt{3}} &\rho^+ & K^{\ast +} \\
\rho^-& \!-\frac{\rho^0}{\sqrt{2}}\!+\!\frac{\omega_8}{\sqrt{6}}\!+\!\frac{\omega_0}{\sqrt{3}}& K^{\ast 0} \\
 K^{\ast -} & \bar{K}^{\ast 0}& \!-\frac{2\omega_8}{\sqrt{6}}\!+\!\frac{\omega_0}{\sqrt{3}} \\
 \end{array}
\!\right)_{\mu\nu}. \nonumber
\end{eqnarray}
The physical $\phi$ and $\omega$ mesons exhibit mixing characterized by an angle $\theta_V$ \cite{Dumm:2009kj,Dai:2013joa}, 
\begin{eqnarray}
\left(\begin{array}{c}
\omega^8\\
\omega^0
\end{array}\right)=\left(\begin{array}{c c}
\cos\theta_V & \sin\theta_V\\
 -\sin\theta_V & \cos\theta_V
\end{array}
\right)\left(\begin{array}{c}
\phi\\
\omega
\end{array}\right)
\end{eqnarray}
The scalars can be filled in the Octet as\footnote{We are aware that the $f_0(500)$ and $f_0(980)$ may not be normal $q\bar{q}$ states, but they should have nonignorable $q\bar{q}$ component \cite{Dai:2011bs, Yao:2020bxx}. Therefore, one can fill these states into the $q\bar{q}$ scalar Octet and apply it to estimate the scalar contributions to the $ee\to4\pi$. } 
\begin{eqnarray}
 S(x) =\left(
 \begin{array}{ccc}
 \frac{\sigma+a_{0}^0}{\sqrt{2}} & a_{0}^+& K_{0}^{*+} \\[3.5mm]
 a_{0}^- & \frac{\sigma-a_{0}^0}{\sqrt{2}}& K_{0}^{*0} \\[3.5mm]
 K_{0}^{*-} & \bar{K}_{0}^{*0} & f_0\\[3.5mm]
\end{array}
\right) \, . \label{eq:Phi}
\end{eqnarray}
The kinetic Lagrangians of the vector and scalar resonance field are given by
\begin{eqnarray}
\label{eq:vkin}
\mathcal{L}_{\mathrm{kin}}^{\mathrm{V}}&=&-\frac{1}{2}\left\langle \nabla^{\lambda}V_{\lambda\mu}\nabla_{\nu}V^{\nu\mu}\right\rangle +\frac{1}{4}M_{V}^{2}\left\langle V_{\mu\nu}V^{\mu\nu}\right\rangle \,,\nonumber\\
\mathcal{L}_{\mathrm{kin}}^{\mathrm{S}}&=&\frac{1}{2}\left\langle \nabla^{\mu}S\nabla_{\mu}S\right\rangle -\frac{1}{2}M_{S}^{2}\left\langle SS\right\rangle \,,\nonumber
\end{eqnarray}
The lowest-order effective Lagrangians for interactions between resonances, goldstone bosons, and photons can be expressed as follows:
\begin{eqnarray}
\mathcal{L}^L_{\rm int}
&=&\frac{F_V}{2\sqrt{2}}\bigl<V_{\mu\nu}f_{+}^{\mu\nu}\bigr>+\frac{iG_V}{\sqrt{2}}\bigl<V_{\mu\nu}u^\mu u^\nu\bigr>\nonumber\\
&+& c_d\bigl<Su_\mu u^\mu\bigr>+c_m\bigl<S\chi_+\bigr> ,
\label{Eq:L:R:2}
\end{eqnarray}
with constriants taken from Refs.~\cite{Ecker:1989yg,Jamin:2001zq},
\begin{equation}
F_VG_V=F^2,\, c_d=c_m=\frac{F}{2}.
\end{equation}
As will be discussed in the next sections, these lowest-order terms are not enough to compensate for the huge discrepancy between the theoretical prediction and experimental data on the cross sections. Therefore, we take into account higher order corrections and the interaction Lagrangians with more than one resonance \cite{Dai:2019lmj},
\begin{eqnarray}
\mathcal{L}^{H}_{\rm int}&=&\lambda_6^V\bigl<V_{\mu\nu}\{f_+^{\mu\nu},\chi_+\}\bigr>+\lambda_{22}^V\bigl<V_{\mu\nu}\nabla_\alpha\nabla^\alpha f_+^{\mu\nu}\bigr> \nonumber\\
&+&\lambda_6^{VV}\bigl<V_{\mu\nu}V^{\mu\nu}\chi_+\bigr>+\lambda_3^{SV}\bigl<\{S,V_{\mu\nu}\}f_+^{\mu\nu}\bigr> \nonumber\\
&+&\lambda^{SVV}\bigl<SV_{\mu\nu}V^{\mu\nu}\bigr>\,. 
\end{eqnarray}
These Lagrangians, together with the lowest order ones of ChPT, Eq.~(\ref{Eq:L;Op2}), construct the interaction Lagrangians involving vectors and scalars that will be applied to analyze the process of $ee\to4\pi$.

\subsection{Formulas for physical observables}
The definition of the momenta in the four Pion production process is as follows: 
\begin{eqnarray}
 e^+(k_1)e^-(k_2)&\rightarrow&\pi^+(p_1)\pi^+(p_2)\pi^-(p_3)\pi^-(p_4)\,,\nonumber\\
 e^+(k_1)e^-(k_2)&\rightarrow&\pi^0(p_1)\pi^0(p_2)\pi^+(p_3)\pi^-(p_4)\,.\nonumber
\end{eqnarray}
The scattering amplitude can be written in terms of 
\begin{equation}
\mathcal{M}=-\frac{4\pi\alpha}{q^2}J_\mu\bar{v}(k_1)\gamma^\mu u(k_2)\,,
\end{equation}
where $q$ is the momentum in the center of the mass frame (c.m.f.), $q=k_1+k_2=\sum_{i=1}^{4}p_i$, and the hadronization vector current $J^\mu$ is 
\begin{equation}
J^\mu\equiv\bigl<\pi(p_1)\pi(p_2)\pi(p_3)\pi(p_4)|\mathcal{J}^\mu_{em}e^{i\mathcal{L}_{QCD}}|0\bigr>\,. \nonumber
\end{equation}
Here, the factor $e^{i\mathcal{L}_{QCD}}$ signifies the hadronization of the electromagnetic current, and it must be assessed amidst the influence of strong interactions. 
The hadronization vector current $J^\mu$ for the 
 $\pi^+\pi^+\pi^-\pi^-$ channel, can be obtained through the $\pi^0\pi^0\pi^+\pi^-$ one by ignoring the isospin breaking effects \cite{Czyz:2000wh,Ecker:2002cw}:
\begin{eqnarray}
J^\mu_c(p_1,p_2,p_3,p_4)
&=&J^\mu_n(p_1,p_3,p_2,p_4)+J^\mu_n(p_1,p_4,p_2,p_3) \nonumber \\
&+&J^\mu_n(p_2,p_3,p_1,p_4)+J^\mu_n(p_2,p_4,p_1,p_3)\,. \nonumber \\
\label{Eq:J;c}
\end{eqnarray}
Here, the subscript \lq $n$' represents the $\pi^0\pi^0\pi^+\pi^-$ production process, and the subscript \lq $c$' is for $\pi^+\pi^+\pi^-\pi^-$ production process.
A convenient choice of the definition of the Dalitz variables is\cite{Unterdorfer:2002zg,Ecker:2002cw}
\begin{eqnarray}
s_{12}&=&(p_1+p_2)^2\,,\nonumber\\
t_i&=&p_i\cdot q,~{\rm with~ i=1,2,3,4}\,,\nonumber\\
\nu&=&\frac{(p_3-p_4)\cdot(p_1-p_2)}{2}\,. \label{Eq:Man;variable}
\end{eqnarray}
There should be a redundant variable according to the relation $q^2=\sum_{i=1}^{4}t_{i}$. Nevertheless, to effectively showcase the symmetries of the form factors, we retain the full set. Also, we will express the form factors using the various scalar products, e.g., $p_i\cdot p_j$ rather than relying solely on $s_{12}$, $t_i$, and $\nu$.

The cross section of $ee\to\pi\pi\pi\pi$ can be expressed as 
\begin{eqnarray}
\sigma\!&\!=\!&\!-\frac{\alpha^2}{3072\pi^4Q^6 S}
\int\!ds_{12}ds_{34}ds_{124}ds_{134}ds_{14}\frac{J^\alpha J_\alpha^*}{\sqrt{-\Delta_4}} \,, \nonumber\\ \label{Eq:cs}
\end{eqnarray}
where one has $Q=\sqrt{q^2}$. $s_{ij}=(p_i+p_j)^2$ and $s_{ijk}=(p_i+p_j+p_k)^2$ are another kind of Mandelstam variables, and one can obtain them from Eq.~(\ref{Eq:Man;variable}), 
\begin{eqnarray}
t_1&=&\frac{s_{12}-s_{34}+s_{134}-m_{\pi}^2}{2}\,, \nonumber\\
 t_2&=&\frac{Q^2-s_{134}+m_{\pi}^2}{2}\,, \nonumber\\ t_3&=&\frac{Q^2-s_{124}+m_{\pi}^2}{2} \,,\nonumber\\
 t_4&=&\frac{-s_{12}+s_{34}+s_{124}-m_{\pi}^2}{2}\,, \nonumber\\
 v&=&Q^2-s_{14}+\frac{s_{134}}{2}+\frac{s_{124}}{2}-\frac{s_{134}}{4}\,.
\label{Eq:Man:v2}
\end{eqnarray}
Notice that the series of variables in Eq.~(\ref{Eq:Man;variable}) are convenient for writing the amplitude in a simplified way, and the other kind in Eq.~(\ref{Eq:Man:v2}) is easy for integrating out the four-body phase space. 
$S$ is the symmetric factor, and one has $S=2$ for the final states $\pi^0\pi^0\pi^+\pi^-$ and $S=4$ for the final states $\pi^+\pi^-\pi^+\pi^-$. 
$\Delta_4$ is a determinant that can be found in Ref.~\cite{Weil:2017knt}, 
\begin{eqnarray}
\Delta_4=\frac{1}{16}\left | \begin{array}{cccc}
2p_1\cdot p_1 &2p_1\cdot p_2 &2p_1\cdot p_3 & 2p_1\cdot p_4 \\
2p_2\cdot p_1 &2p_2\cdot p_2 &2p_2\cdot p_3 & 2p_2\cdot p_4 \\
2p_3\cdot p_1 &2p_3\cdot p_2 &2p_3\cdot p_3 & 2p_3\cdot p_4 \\
2p_4\cdot p_1 &2p_4\cdot p_2 &2p_4\cdot p_3 & 2p_4\cdot p_4 
\end{array} \right | \,. \nonumber
\end{eqnarray} 
The upper and lower limits of the integration can be found in the Appendix \ref{appendix:ps}.
Notice that the mass of the electron is ignored.

\section{Amplitudes from ChEFT}
As discussed before, we have two ways to perform the analysis of $e^+e^-\to \pi\pi\pi\pi$. One is from ChPT, and the other is from RChT. In the latter, the $\mathcal{O}(p^2)$ Lagrangians of ChPT will also be included. 

\subsection{Amplitudes at leading order}
\begin{figure}[!htb]
\centering
\includegraphics[width=0.35\textwidth]{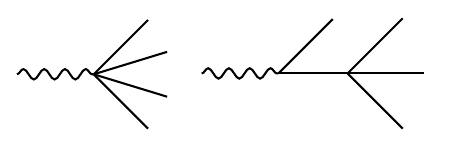} 
\caption{The LO Feynman diagrams for $\gamma^*\rightarrow 4\pi$. 
The solid lines represent pions while the wavy line symbolizes a virtual photon, and so on for the following diagrams.}
\label{Fig:Op2}
\end{figure}

The hadronization vector current form factor of $\gamma^*\rightarrow 4\pi$ from ChPT can be written as 
\begin{equation}
J^\mu_{\rm ChPT}=J^\mu_{(2)}+J^\mu_{(4)} \,, \label{Eq:J:ChPT}
 \end{equation}
where the subscripts \lq 2,4' are for chiral counting. 
Note that this form factor is for the process of $e^+e^-\to\pi^0\pi^0\pi^+\pi^-$. See Eq.~(\ref{Eq:J;c}).We ignore the subscript \lq n' of $J_\mu$ for simplicity, and so on for discussions below if there is no specification. 
The first part of Eq.(\ref{Eq:J:ChPT}) is calculated from the tree diagrams, which are illustrated in Fig.\ref{Fig:Op2}. 
The relevant chiral Lagrangians are of $\mathcal{O}(p^2)$ $SU(3)$ ChPT ones. See Eq.~(\ref{Eq:L;Op2}). 
The form factor for the process of $e^+e^-\to\pi^0\pi^0\pi^+\pi^-$ at $\mathcal{O}(p^2)$ is given as 
\begin{equation}
 J^\mu_{(2)}(p_1,p_2,p_3,p_4)=\frac{s_{12}-m_\pi^2}{F_\pi^2}\biggl(\frac{2p_3^\mu}{2t_3-q^2}-\frac{2p_4^\mu}{2t_4-q^2}\biggr)\ .\label{Eq:J2}
\end{equation}
Notice that vector current for $e^+e^-\to\pi^+\pi^+\pi^-\pi^-$ can be obtained through Eq.~(\ref{Eq:J;c}).

\subsection{ChPT amplitudes at $\mathcal{O}(p^4)$}
For amplitudes at NLO, there are typically two components: the one from one-loop diagrams that use vertices from the $\mathcal{O}(p^2)$ chiral effective Lagrangians as given in Eq.~(\ref{Eq:L;Op2}), while the other component originates from the tree-level diagrams generated by the $\mathcal{O}(p^4)$ Lagrangian in Eq.~(\ref{Eq:L;Op4}).
\begin{figure}[!htb]
\centering
\includegraphics[width=0.45\textwidth]{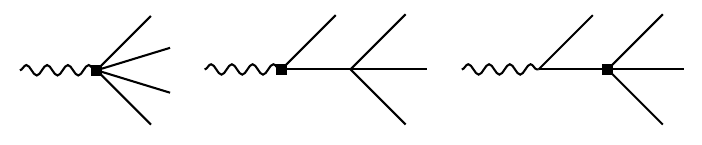} 
\includegraphics[width=0.45\textwidth]{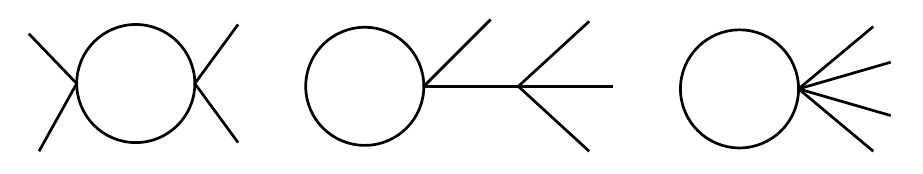}
\caption{The Feynman diagrams for the NLO form factors of $\gamma^*\to 4\pi$. The vertices of the filled square are derived from the $\mathcal{O}(p^4)$ effective Lagrangians. The bottom graphs are one-loop diagrams, where the virtual photon can be attached to all allowed lines and vertices. Diagrams illustrating wave function renormalization are not shown.} \label{Fig:Feynman;Op4}
\end{figure}
The Feynman diagrams for NLO form factors are shown in Fig.~\ref{Fig:Feynman;Op4}. In the first row, the vertex (filled square) is from $\mathcal{O}(p^4)$ ChPT Lagrangians. In the second row, a virtual photon should be included whenever possible. 
The hadronization vector current form factor (for Solution I)  can be expressed as 
\begin{equation}
 J^\mu_{(4)}=J^\mu_{(4)tree}+J^\mu_{(4)loop} \,.
\end{equation}
The two parts of the form factors mentioned above are detailed in Appendix \ref{appendix:M}. 

\subsection{RChT amplitudes}
\label{section6}
\begin{figure}[!htb]
\centering
\includegraphics[width=0.45\textwidth]{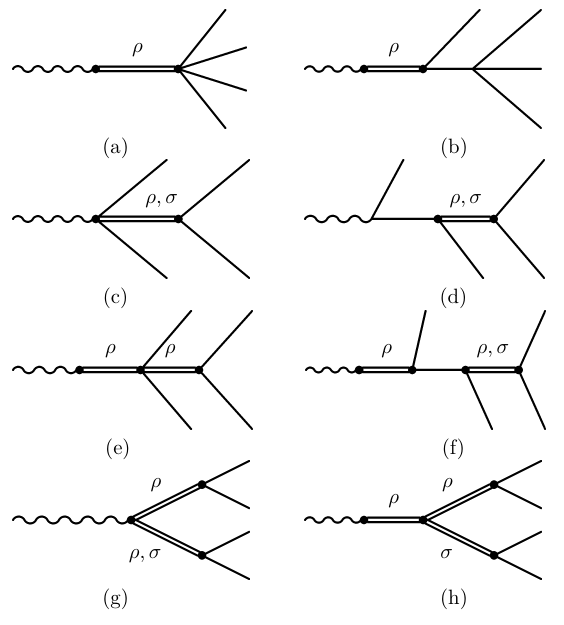}
\caption{In the diagrammatic notation, the $\rho$,$\sigma/f_0(500)$ and $f_0(980)$ mesons are represented by double parallel lines.}
\label{Fig:Feynman;RChT}
\end{figure}

 The Feynman diagrams, including exchanges with resonances, are shown in Fig.~\ref{Fig:Feynman;RChT}. 
Note that the lowest order result from ChPT should be included without the double counting problem. See Fig.\ref{Fig:Op2}. 
The reason is that by integrating out the resonance in RChT Lagrangians, it would contribute operators of pseudoscalars with chiral counting no less than $\mathcal{O}(p^4)$. Hence, the interaction Lagrangians in Eq.~(\ref{Eq:L:R:2}) will be at least $\mathcal{O}(p^4)$ in this way. 
Finally, the form factor of the hadronization vector current from RChT (for Solution II) is 
\begin{equation}
 J_{\rm RChT}^\mu=J^\mu_{(2)}+J^\mu_{R}\,,
\end{equation}
where $J^\mu_{(2)}$ is from Eq.~(\ref{Eq:J2}) and $J^\mu_{R}$ can be found in Appendix \ref{appendix:RChT}. 
Notice that the diagrams with $\rho$, $\rho\rho$, $\sigma$ meson exchanges have been considered in Ref.~\cite{Ecker:2002cw}, including the first six diagrams in Fig.\ref{Fig:Feynman;RChT}. However, these contributions are still not enough to describe the cross sections in the low-energy region. 
Also, they have taken into account the $f_2(1270)$, $a_1(1260)$ and $\omega(782)$. The first two resonances are too far away from the low energy region we focused on, and the $\omega$ has a rather small width, resulting in a negligible contribution to the energy region $E_{c.m.}\leq~0.6$~GeV. Indeed, we have also considered the contributions from $f_2(1270)$, $a_1(1260)$ and $\omega(782)$; their contributions are much smaller than the other mesons mentioned above.

\begin{figure}[!htb]
\centering
\includegraphics[width=0.45\textwidth]{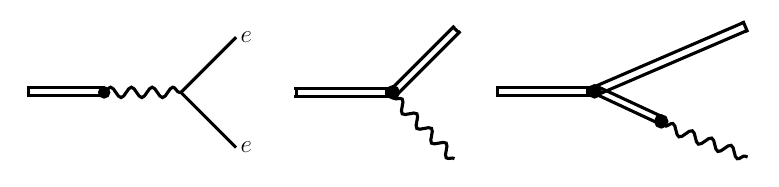}
\caption{ The first graph is for the decay process of $\rho,\omega,\phi\to e^+e^-$. The last two graphs are for the decay width of $\phi\to a_0^0\!+\!\gamma, f_0\!+\!\gamma$. }
\label{Fig:width}
\end{figure}

To fix the unknown couplings, we also calculate the decay widths of $\rho,\omega,\phi\to e^+e^-$ and $\phi\to a_0(980) \gamma, f_0(980)\gamma$. 
The Feynman diagrams are shown in Fig.~\ref{Fig:width}. 
We do not list more processes and decay widths, as these are enough to fix the parameters. 

As outlined in the previous section, in Solution III we incorporate the one-loop ChPT terms into the RChT framework. The corresponding hadronization vector current is given by
\begin{equation}
J^\mu_{\rm Sol.~III}=J^\mu_{(2)}+J^\mu_{(4){\rm loop}}+J^\mu_{R}\,. \nonumber \label{Eq:J:TMI}
\end{equation}

\section{Results and Discussions} 
As we have stated in the previous sections, there are two solutions, Solution I is from ChPT, and Solution II is from RChT. In the first solution, the LECs have been fixed by other groups \cite{Bijnens:2014lea}, and we apply them directly. For Sol.~II, the unknown couplings need to be fixed, and we apply MINUIT \cite{James:1975} to fit the vector masses and the decay widths mentioned above. For Solution III, all the parameters are taken from Solutions I and II.

The unknown couplings are $\lambda_6^V$, $\lambda_{22}^V$, $\lambda_6^{VV}$, $\lambda_3^{SV}$ and $\lambda^{SVV}$.
For the value of $\lambda_6^{VV}$, it can be estimated by considering the physical masses between $M_\phi$ and $M_\rho\backsimeq M_\omega$. 
One can expand the Lagrangian of $\lambda_6^{VV}\bigl<V_{\mu\nu}V^{\mu\nu}\chi_+\bigr>$ as $(4\lambda_6^{VV}m_k^2-2\lambda_6^{VV}m_\pi^2)\phi^2$, which should be added to the kinetic Lagrangian term of $\frac{1}{4}M_V^2\bigl<V_{\mu\nu}V^{\mu\nu}\bigr>\supset \frac{1}{4}M_V^2\phi^2$.
Finally one has $\frac{1}{4}M_V^2\!+\!4\lambda_6^{VV}m_k^2\!-\!2\lambda_6^{VV}m_\pi^2\!\approx\!\frac{m_\phi^2}{4}$, resulting in$\lambda_6^{VV}\!\approx\!0.116$. Notice that one has $M_V=M_\rho$.
For $\lambda_6^V$ and $\lambda_{22}^V$, we can fit the decay widths of $\rho,\omega,\phi\rightarrow e^+\!+\!e^-$ to determine their values.
For $\lambda_3^{SV}$ and $\lambda^{SVV}$, they can be fixed by fitting to the decay widths of $\phi\rightarrow a_0^0\!+\!\gamma, f_0\!+\!\gamma$.
The remain parameters, $F_V$ and $\theta_V$ are taken from Fit.~1 of Ref.~\cite{Dai:2013joa}, $F_V=0.142$~GeV and $\theta_V=39.45^\circ$. 

\subsection{Fit to the decay widths}
This subsection is for Sol .~II, as one needs to fit the decay widths to fix the unknown couplings of RChT. 
For the decay widths of $\phi\rightarrow a_0\!+\!\gamma, f_0\!+\!\gamma$, the masses of the $a_0$ and $f_0$ mesons are very close to the mass of the $\phi$ meson, and hence the phase space would significantly affect the widths. For the mass of the $f_0$, we apply the pole mass obtained in a comprehensive amplitude analysis \cite{Dai:2014lza,Dai:2014zta}, $M_{f_0}=0.998$~GeV. For the mass of the $a_0$, its mass is less definitive, and we set it as a parameter that needs to be fixed by fitting the data. As can be found, finally we have $M_{a_0}=0.9769\pm0.0018$~GeV, which is consistent with the value of PDG \cite{ParticleDataGroup:2022pth} within the uncertainty. 
We list the fit results in Table \ref{Tab:width}, where the relevant Feynman diagrams are shown in Fig.~\ref{Fig:width}.
\begin{table}[h]
\centering
\begin{tabular}{c c c}
\hline
\hline
\rule[-0.2cm]{0cm}{6mm}Parameter & fit & PDG \\
\hline
$\lambda_6^V({\rm GeV}^{-1})$ & $-0.0636\pm0.0009 $&\quad\\
$\lambda_{22}^V({\rm GeV}^{-1})$ & $-0.0168\pm0.0007$&\quad\\
$\lambda_{6}^{VV}({\rm GeV}^{0})$ & 0.116&\quad\\
$\lambda_{3}^{SV}({\rm GeV}^{0})$ & $-3.7707\pm0.1307$&\quad\\
$\lambda^{SVV}({\rm GeV}^{1})$&$6.3030\pm1.0550$&\quad\\
$M_{a_0^0}({\rm GeV})$&$0.9769\pm0.0018$&$0.98\pm0.02$\\
$M_{f^0}({\rm GeV})$& 0.998\cite{Dai:2014zta}&$0.99\pm0.02$\\
$\Gamma_{\phi\rightarrow a_0\gamma}[10^{-7}{\rm GeV}]$ 
&$3.0886\pm0.1302$&$3.229\pm0.265$\\
$\Gamma_{\phi\rightarrow f_0\gamma}[10^{-6}{\rm GeV}]$ 
&$1.3283\pm0.0374$&$1.368\pm0.085$\\
$\Gamma_{\rho\rightarrow e^+e^-}[10^{-6}{\rm GeV}]$ 
&$7.023\pm0.050$&$7.038\pm0.112$\\
$\Gamma_{\omega\rightarrow e^+e^-}[10^{-7}{\rm GeV}]$ 
&$6.106\pm0.053$&$6.406\pm0.287$\\
$\Gamma_{\phi\rightarrow e^+e^-}[10^{-6}{\rm GeV}]$ 
& $1.180\pm0.027$ & $1.266\pm0.018$\\
\hline
\hline
\end{tabular}
\caption{The fit results of fitting to the decay widths.}
\label{Tab:width}
\end{table}

We apply MINUIT \cite{James:1975} to fix the unknown couplings. 
The uncertainties of the parameters are given by MINUIT, and those of the decay widths are estimated from bootstrap \cite{efron1992bootstrap} within $1\sigma$, which is performed by varying the parameters with the MINUIT uncertainty multiplying a normal distribution function. 
The $\chi^2/N$ is given as $2.30$, where $N$ is the number of data points. Though it is a bit far away from 1, the fit quality is not bad, as our results are rather close to those of PDG.
Indeed, all the processes have quite small $\chi^2$ except for the one of $\Gamma_{\phi\rightarrow e^+e^-}$ ($\chi^2=10.67$). 
The reason may be that our theory does not take into account the final state interactions of $V\to PP$, with $P$ the pion or kaon.
Nevertheless, our width of $\Gamma_{\phi\rightarrow e^+e^-}$ is still compatible with that of the PDG, considering the uncertainties of both. The fit result of all the decay widths implies that all the parameters of RChT are well fixed. 
In the following subsection, we will input these parameters into the amplitudes and calculate the cross sections of $e^+e^-\to\pi^+\pi^+\pi^-\pi^-$ and $e^+e^-\to\pi^0\pi^0\pi^+\pi^-$.

\subsection{Prediction on the cross section}
\begin{figure}[!htb] 
\centering
\includegraphics[width=0.48\textwidth]{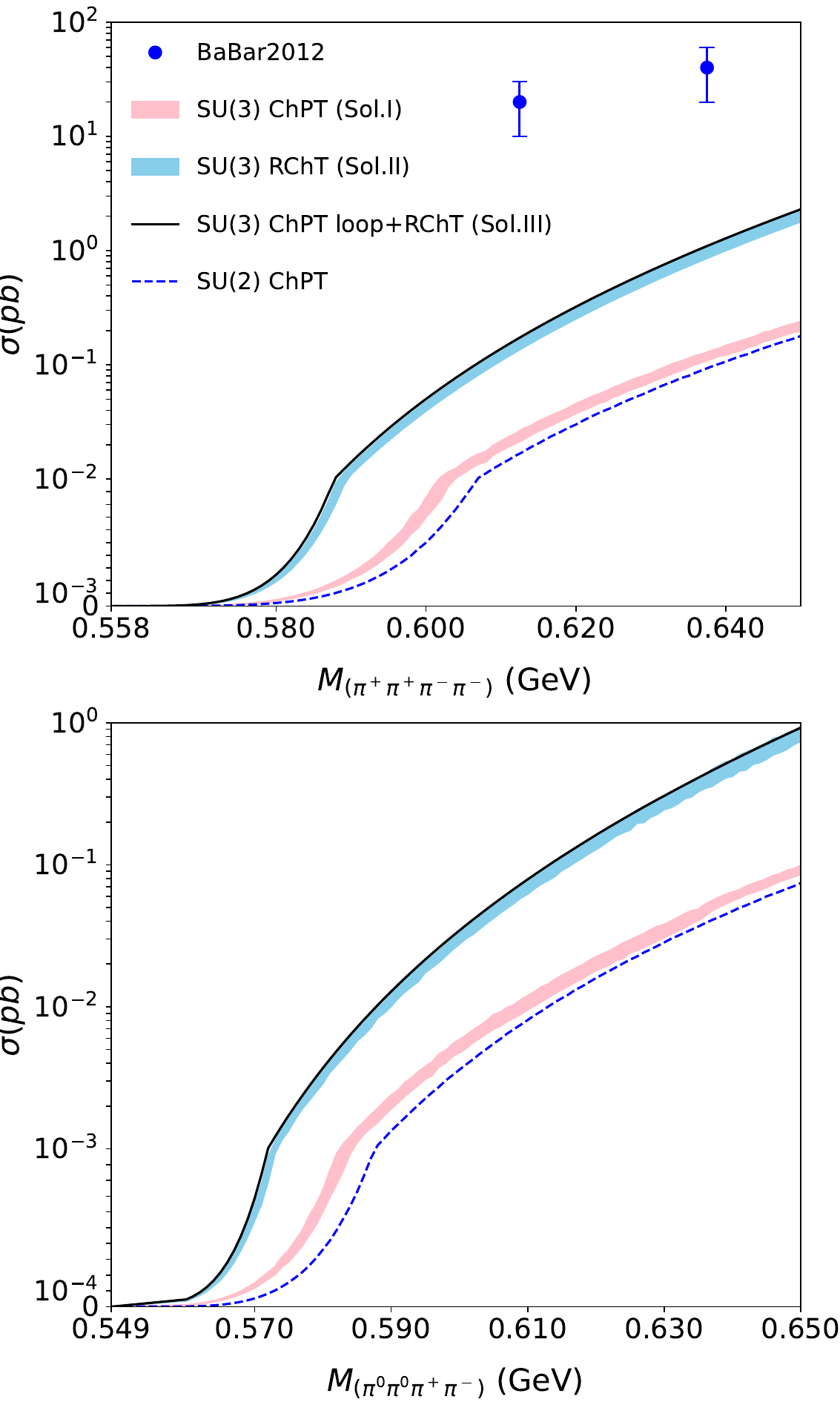}
\caption{The predictions of cross section from $SU(3)$ ChPT and RChT, shown as the pink and cyan bands. The $SU(2)$ ChPT result, shown as blue dashed lines, is also plotted for comparison. }
\label{Fig:cs}
\end{figure}

The cross sections for electron-positron annihilating into four pions can be obtained through Eq.~(\ref{Eq:cs}). It would be safe to restrict our analysis up to 0.6~GeV, as it is roughly 175~MeV from the central value of the $\rho$ resonance. Nevertheless, there are two experimental data points for the total cross-section of $e^+e^-\to\pi^+\pi^+\pi^-\pi^-$ in the energy region of [0.60, 0.65]~GeV, which helps people to judge the consistency between the data and our prediction. Thus, we plot our cross sections up to 0.65~GeV. Notice that it is still 125~MeV away from the center mass of the $\rho$ resonance. 
The cross sections of our solutions are shown in Fig.~\ref{Fig:cs}. 

For Sol.~I, the $SU(3)$ ChPT result up to NLO is shown as the pink band. The error bands are estimated again from a bootstrap-like method, varying the LECs within their uncertainty~\cite{Bijnens:2014lea} and choosing the deviation of $1\sigma$ from the central value as the error.
We also plot the $SU(2)$ ChPT result as the blue dashed lines, which has already been estimated in Ref.~\cite{Ecker:2002cw} before. As can be found, the $SU(2)$ and $SU(3)$ ChPT results are compatible with each other. This is not a surprise as both should work well in the low-energy region. Also, the difference between them is distinct. This is caused by the fact that the $SU(3)$ result includes the kaon loops, etc..
Meanwhile, both are approximately two orders of magnitude smaller than the experimental data of $e^+e^-\to\pi^+\pi^+\pi^-\pi^-$. See the two data points (blue filled circles) in the energy region of [0.6, 0.65]~GeV in the first graph of Fig.\ref{Fig:cs}. It reveals that resonances, especially $\rho(770)$ and $\sigma$, should be included to compensate for the significant discrepancy.

For Sol.~II, we apply RChT to clarify the contribution of the resonances. The result is shown as the cyan bands in Fig.\ref{Fig:cs}, which are again estimated by varying the couplings within its uncertainty (given by MINUIT), and the deviation of $1\sigma$ from the central value is extracted as the error. 
As can be found, the RChT results are several times larger than those of $SU(3)$ ChPT.

For Sols.~III, it has modest differences with Sol.~II. Hence, it is resented by lines instead of bands. This close agreement stems from the fact that the $\mathcal{O}(p^4)$ ChPT contributions, from both one-loop corrections and tree diagrams with LECs, are substantially outweighed by the resonance terms. In the following sections, we will focus mainly on Sols.~I and II. 
    
From the results at $Q=0.6125$~GeV, one can compare the ChEFT results with the data. Ours are roughly three to four orders smaller than that of the experiment, $0.02\pm0.01$ nb \cite{BaBar:2012sxt}. However, one should also be careful with the large uncertainty of the experiment.
Therefore, we urge future measurements of the four-pion production, especially in the energy region below 0.6~GeV.

\begin{figure}[!htb] 
\centering
\includegraphics[width=0.48\textwidth]{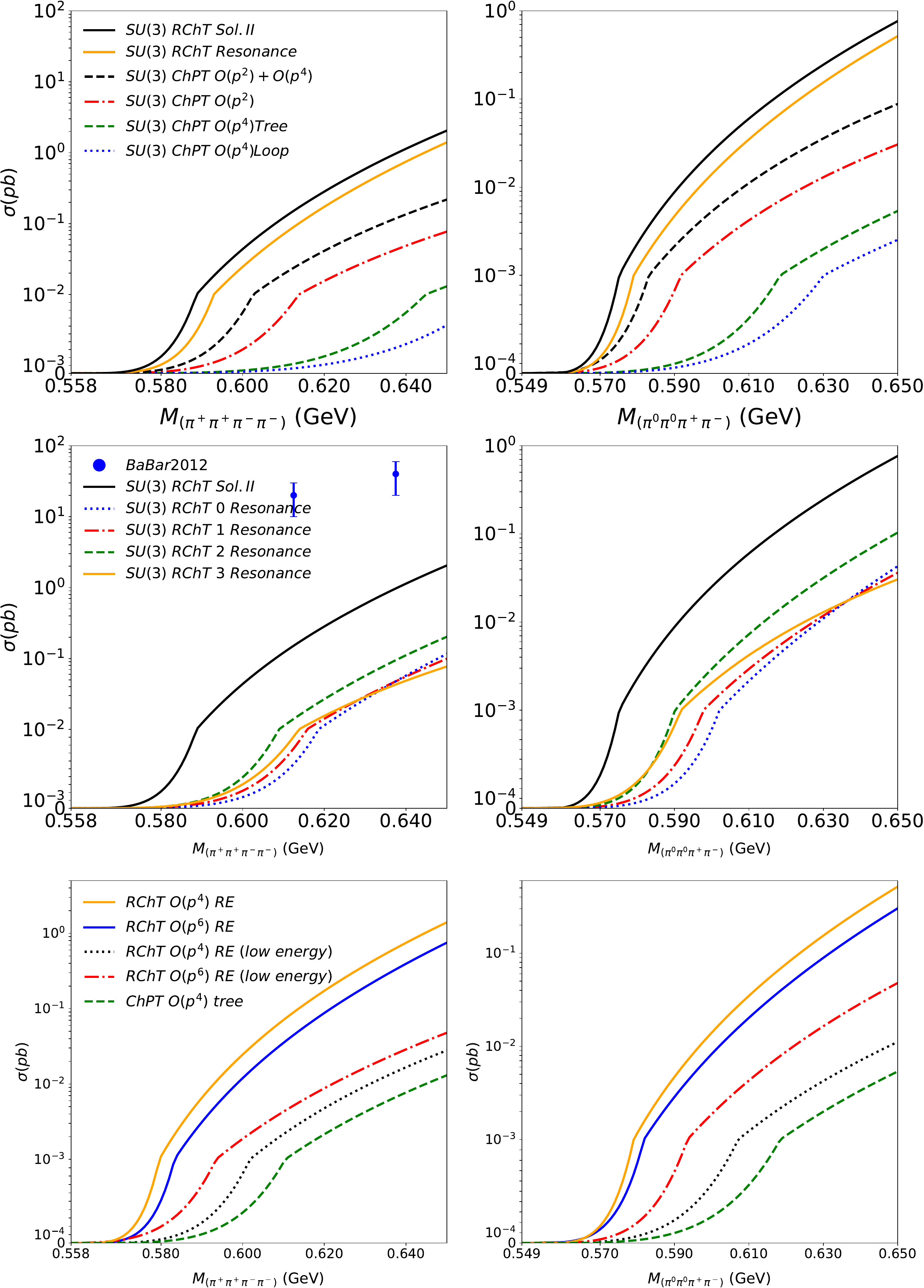}
\caption{The individual contributions to the cross sections from different parts of the amplitude. For the bottom graphs, the label 'low energy' indicates that the momentum in the resonance propagator has been removed. \lq RE' represents resonance exchange.}
\label{Fig:ind} 
\end{figure}
The individual contributions to the cross sections from the various components of the amplitude are shown in Fig.~\ref{Fig:ind}, comprising the ChPT $\mathcal{O}(p^2)$ tree diagrams, ChPT $\mathcal{O}(p^4)$ tree diagrams, ChPT $\mathcal{O}(p^4)$ one-loop diagrams, and RChT tree diagrams with resonances. Sol.~III bear a strong resemblance to Sol.~II, with all their key components being encompassed by the latter. Thus, our following analysis is confined to the first two solutions. 
The graphs on the left side are for $e^+e^-\to\pi^+\pi^+\pi^-\pi^-$, and the ones on the right side are for the $e^+e^-\to\pi^0\pi^0\pi^+\pi^-$. 

The upper panels show the individual components from ChPT (Sol.~I) and RChT (Sol.~II), respectively, while the lower panels show the contributions from different combinations of the resonances within RChT (Sol. II).
For the grpahs in the upper panel, the contribution from tree diagrams of $\mathcal{O}(p^2)$ effective Lagrangians, shown as the red dash-dotted lines, dominates the contributions of ChPT. The contributions from tree diagrams of $\mathcal{O}(p^4)$ effective Lagrangians (Olive dashed lines) and the one from one-loop diagrams of $\mathcal{O}(p^2)$ effective Lagrangians (Blue dotted lines) are roughly one order smaller. Note that the latter is smaller than the former. This observation is compatible with the chiral counting. Moreover, all the contributions from ChPT terms are much smaller than that of the resonances, shown as the orange solid line. Accordingly, no notable discrepancies are observed among Sols.~II and III.

For the graphs in the middle panel, the contributions from diagrams with one, two, three, or no resonances are roughly in the same order. See the red dash-dotted, olive dashed, orange solid and blue dotted lines, respectively. The relevant Feynman diagram without resonance is calculated from the effective Lagrangians $\mathcal{O}(p^2)$ of ChPT, as shown in Fig.~\ref{Fig:Op2}. The Feynman diagrams with several resonances are indicated in Fig.~\ref{Fig:Feynman;RChT}. 
Notice that the RChT result without resonance is the same as that of the ChPT tree diagrams $\mathcal{O}(p^2)$. So it is not strange that the total results of RChT are roughly one order larger than those of ChPT. 

For the graphs in the bottom panel, it can be seen that the “$SU(3)$ RChT resonance exchange (RE) $\mathcal{O}(p^4)$ (low-energy)” contribution (black dotted lines) is close to that of the ChPT $\mathcal{O}(p^4)$ tree diagrams (LEC terms), whereas the “$SU(3)$ RChT RE $\mathcal{O}(p^6)$ (low-energy)” contribution (red dash-dotted lines) is larger. The reason is that the $\mathcal{O}(p^4)$ RE originates from operators such as $\frac{F_V}{2\sqrt{2}} \langle V_{\mu\nu}f_{+}^{\mu\nu}\rangle$ and $\frac{iG_V}{2\sqrt{2}} \langle V_{\mu\nu}u^\mu u^\nu\rangle$, which contribute at leading order to the $\mathcal{O}(p^4)$ ChPT LECs after integrating out the resonances. In contrast, the term $\lambda^{SVV} \langle S V_{\mu\nu} V^{\mu\nu}\rangle$ contributes at leading order to the $\mathcal{O}(p^6)$ ChPT LECs.
Nevertheless, a noticeable difference remains between the “RChT RE $\mathcal{O}(p^4)$ (low-energy)” and the “ChPT $\mathcal{O}(p^4)$ LEC” results. This may be attributed to the following factors: the considered energy region (e.g., 0.52–0.6 GeV) is too close to the resonance region, particularly that of the $\sigma$ resonance, and it does not lie within the genuinely low-energy regime. Furthermore, the LECs we employ are not the ones obtained by matching RChT to ChPT in the low-energy region \cite{Pich:2002xy}, but they are taken from phenomenological analysis~\cite{Bijnens:2014lea}. Additionally, the procedure of removing the momentum dependence in the propagator may only provide a simplified comparison between ChPT and RChT in the low-energy region.

Further, the full REs, where the momentum dependence in the propagators is retained, yield contributions that are considerably larger than those from the “low-energy” resonance exchanges. This helps clarify why the cross sections for $4\pi$ production differ so significantly between the ChPT and RChT approaches. Higher-order contributions, such as those at $\mathcal{O}(p^6)$ and beyond, could account for the discrepancies between the ChPT and RChT results.

Moreover, one would notice that the contribution with no resonance has the most significant contributions in the low energy region, e.g., $Q\le 0.58$~GeV, but has fewer contributions in the higher energy region. This is compatible with our conclusion: the resonance contributions can not be ignored, and they should become more and more important in the higher energy region.

\subsection{The contribution to $(g-2)_\mu$}
The formalism for the leading-order (LO) correction to the anomalous magnetic moment of the muon, denoted as $a_\mu\equiv(g-2)_\mu$, can be derived from the cross sections of electron-positron annihilation into hadrons, utilizing the optical theorem and principles of analyticity. One has \cite{Brodsky:1967sr,Lautrup:1968tdb}
\begin{equation}
 a_\mu^{HVP,LO}=\frac{\alpha_e^2(0)}{3\pi^2}\int_{{\rm th}}^\infty ds\frac{\hat{K}(q^2)}{q^2}R_h(q^2).
\end{equation}
where $\alpha_e(0)=\frac{e^2}{4\pi}$ denotes the electromagnetic fine-structure constant, with \lq th' indicating the threshold. $\hat{K}(q)$ is the kernel function defined as follows \cite{Aoyama:2020ynm,Brodsky:1967sr,Lautrup:1968tdb},
\begin{equation}
\begin{aligned}
\hat{K}(q^2)=\biggl[\biggl(\ln(1+&x)-x+\frac{x^2}{2}\biggr)\frac{(1+x^2)(1+x)^2}{x^2}\\
&+\frac{x^2}{2}(2-x^2)+\frac{1+x}{1-x}x^2\ln x\biggr]
\end{aligned}
\end{equation}
with
\begin{equation}
 x=\frac{1-\rho_\mu(q^2)}{1+\rho_\mu(q^2)},\qquad \rho_\mu(q^2)=\sqrt{1-\frac{4m_\mu^2}{q^2}}
\end{equation}
The hadronic R-ratio is
\begin{equation}
 R_h(q^2)=\frac{3q^2}{4\pi\alpha_e^2(q^2)}\sigma\bigl(e^+e^-\rightarrow hadrons\bigr),
\end{equation}
which we applied the values given by Ref.~\cite{Wang:2023njt}.

The prediction of $a_\mu$ from the process of $e^+e^-\rightarrow 4\pi$ at LO is presented in Table \ref{Tab:g-2}. 
\begin{table}[h]
\centering
\begin{tabular}{c c c}
\hline
\hline 
\rule[-0.2cm]{0cm}{6mm}& $SU(3)~{\rm ChPT}$  & $SU(3)~{\rm RChT}$\\
\hline
$a_\mu^{2\pi^+2\pi^-}[q\leq0.6 {\rm GeV}]$ & $0.128\pm0.014$ & $0.680\pm0.062$ \\
$a_\mu^{2\pi^0\pi^+\pi^-}[q\leq0.6{\rm GeV}]$ & $0.112\pm0.012$ & $0.597\pm0.058$ \\
$a_\mu^{2\pi^+2\pi^-}[q\leq0.65{\rm GeV}]$ & $7.79\pm0.73$ & $62.54\pm7.23$ \\
$a_\mu^{2\pi^0\pi^+\pi^-}[q\leq0.65{\rm GeV}]$ & $3.57\pm0.32$ & $28.35\pm3.44$\\
\hline
\hline
\end{tabular} 
\caption{The contributions to the $a_{\mu}$ from electron-positron annihilation into four pions. All the values should be multiplied by $10^{-16}$. The results of RChT are taken from Sol.~II.  }
\label{Tab:g-2}
\end{table}
The uncertainties are estimated similarly to the Bootstrap method \cite{efron1992bootstrap}, where we varied the parameters within their uncertainties randomly, and the errors of the $a_\mu$ are chosen as $1\sigma$ away from the central values.
The contribution from the $\pi^+\pi^+\pi^-\pi^-$ channel to $a_\mu$ up to 0.6 GeV are $a_\mu=(0.128\pm0.014)\times10^{-16}$ for ChPT (Sol.~I) and $a_\mu=(0.680\pm0.062)\times10^{-16}$ for RChT (Sol.~II), respectively. 
The contributions from the $\pi^0\pi^0\pi^+\pi^-$ channel up to 0.6 GeV are given as $a_\mu=(0.112\pm0.012)\times10^{-16}$ for ChPT  (Sol.~I) and $a_\mu=(0.597\pm0.058)\times10^{-16}$ for RChT (Sol.~II). 
For the reader's convenience, we also give the contribution up to 0.65 GeV. They are one to two orders larger than the ones up to 0.6~GeV. See Table \ref{Tab:g-2}. 
This is consistent with the observation on the cross section: there is a large enhancement of the cross section caused by the resonances. 
This reveals that the higher order corrections of ChEFT and the FSI of four pions are non-ignorable. Especially, the resonance contribution, not only from the $\rho(770)$ but also the $f_0(500)$, should be taken into account.

\section{Conclusion and Summary}
In this paper, we analyze the processes of electron-positron annihilating into $\pi^+\pi^+\pi^-\pi^-$ and $\pi^0\pi^0\pi^+\pi^-$ final states within ChEFT. With $SU(3)$ ChPT, we predict the scattering cross sections for these channels from threshold up to 0.6 GeV. We extend our predictions to the energy region of 0.6-0.65~GeV, which are two to three orders smaller compared with the two data points of $e^+e^-\to\pi^+\pi^+\pi^-\pi^-$. Further, we apply RChT to include the contributions of the resonances, $\rho$, $\sigma$, etc.. 
The prediction of the cross section is enlarged for one order, but it is still one to two orders smaller than the two data points. 
Considering that the two data points are with poor statistics, we urge the experiments to measure the cross section of four pion production in the low energy region, which would be very helpful for studying the non-perturbative property of the strong interactions. 
Also, we estimate the contributions of these two channels to $(g-2)_\mu$ in the energy range from threshold to 0.6 or 0.65 GeV. 
For ChPT, the contributions from $\pi^+\pi^+\pi^-\pi^-$ channel to $a_\mu$ is $(0.128\pm0.014)\times10^{-16}$ up to 0.6 GeV, and that from $\pi^0\pi^0\pi^+\pi^-$ channel is $(0.112\pm0.012)\times10^{-16}$. 
For RChT, the contributions from $\pi^+\pi^+\pi^-\pi^-$ channel (up to 0.6 GeV) is $(0.680\pm0.062)\times10^{-16}$, and that from $\pi^0\pi^0\pi^+\pi^-$ channel is $a_\mu=(0.597\pm0.058)\times10^{-16}$, respectively. 
The ones up to 0.65~GeV are roughly two orders larger, as shown in Table \ref{Tab:g-2}. 
The higher order corrections of ChEFT, final state interactions, and the light resonances are essential to understanding these processes, and we urge further experimental measurements around the threshold to give more guidance and constraint for theory analysis.

\section*{Acknowledgments}
This work is supported by the National Natural Science Foundation of China (NSFC) with Grants No.12322502, 12335002, Joint Large Scale Scientific Facility Funds of the NSFC and Chinese Academy of Sciences (CAS) under contract No.U1932110, Hunan Provincial Natural Science Foundation with Grant No.2024JJ3004, and Fundamental Research Funds for the central universities.

\appendix

\section{Four-body phase space}\label{appendix:ps}
In this section, we discuss how to deal with the four-body phase space.
The cross section of the process of $e^+e^-\to 4\pi$ can be written as 
\begin{equation}
\sigma=\frac{64\pi^6\alpha^2 }{S q^4\sqrt{(k_1\cdot k_2)}}[k_1^\mu k_2^\nu+k_1^\nu k_2^\mu-g^{\mu\nu}k_1\!\cdot k_2]\mathcal{T}_{\mu\nu}(q^2), \nonumber
\end{equation}
where the electron mass is ignored, and one defines
\begin{equation}
\mathcal{T}_{\mu\nu}(q^{2})=\int ds_{12}ds_{34}ds_{124}ds_{134}ds_{14}
\hat{A}\cdot J_{\mu}J_{\nu}^{*}\,, \nonumber
\end{equation}
with 
\begin{eqnarray}
\hat{A}&=&\frac{1}{(2\pi)^{12}}\frac{\pi^2}{32q^2}\frac{1}{\sqrt{-\Delta_4}}. \nonumber
\end{eqnarray}
Following the Lorentz structure decomposition and $Ward\ Identity$, the tensor $\mathcal{T}_{\mu\nu}(q^2)$ after the integration can be re-written as 
\begin{equation}
\mathcal{T}_{\mu\nu}(q^{2})=(g_{\mu\nu}q^{2}-q_{\mu}q_{\nu})\frac{g^{\alpha\beta}\mathcal{T}_{\alpha\beta}(q^{2})}{3q^{2}}\,. \label{Eq:T;tensor}
\end{equation}
The terms proportional to $q^\mu$ must be restored in the hadronized vector current $J^\mu$, as they are non-vanishing in the scalar function $g^{\alpha\beta}J_\alpha J_\beta$. Pursuant to the Ward identity, the full current is given by 
\begin{eqnarray}
J_{\rm full}^\mu= J^\mu+\frac{q^\mu q\cdot J}{q^2},.
\end{eqnarray}
It should be noted that the expressions for the current form factor in later sections omit these $q^\mu$ terms.

As a result, the formulas for the cross section can be written as 
\begin{eqnarray}
\sigma=\!\!-\frac{\alpha^2}{3072\pi^4q^6 S}\!
\int\!ds_{12}ds_{34}ds_{124}ds_{134}ds_{14}\frac{J^\alpha J_\alpha^*}{\sqrt{-\Delta_4}} \,.\nonumber\\
\end{eqnarray}
This is Eq.(\ref{Eq:cs}). The integration limits of $s_{12}$, $s_{34}$, $s_{124}$, $s_{134}$ and $s_{14}$ are given in Ref.~\cite{Weil:2017knt}, 
\begin{widetext}
\begin{eqnarray}
s_{12}^{+}&=&(q-2m_\pi)^2,\ \ s_{12}^-=4m_\pi^2,\ \ s_{34}^+=(q-\sqrt{s_{12}})^2,\ \ s_{34}^-=4m_\pi^2, \nonumber\\
s_{124}^{\pm}&=&\frac{1}{2s_{34}}[-s_{34}^{2}+s_{34}(s_{12}+m_{4}^{2}+m_{3}^{2}+q^{2})-(s_{12}-q^{2})(m_{4}^{2}-m_{3}^{2})]\pm\frac{1}{2s_{34}}\lambda^{\frac{1}{2}}(s_{34},s_{12},q^2) \lambda^{\frac{1}{2}}(s_{34},m_3^2,m_4^2)\,,\nonumber\\
s_{134}^{\pm}&=&\frac{1}{2s_{12}}[-s_{{12}}^{2}+s_{12}(s_{34}+m_{1}^{2}+m_{2}^{2}+q^{2})-(s_{34}-q^{2})(m_{1}^{2}-m_{2}^{2})]\pm\frac{1}{2s_{12}}\lambda^{\frac{1}{2}}(s_{12},s_{34},q^2)\lambda^{\frac{1}{2}}(s_{12},m_2^2,m_1^2)\,,\nonumber\\
s_{14}^{\pm}&=&\frac{-1}{\lambda(s_{34},s_{12},q^{2})}
\left|
\begin{array}{ccc}
 2s_{34}&s_{34}+s_{12}-q^{2}&s_{34}-s_{134}+m_{1}^{2}\\
 s_{34}+s_{12}-q^{2}&2s_{12}&s_{12}-m_{2}^{2}+m_{1}^{2}\\
 s_{34}-m_{3}^{2}+m_{4}^{2}&s_{12}-s_{124}+m_{4}^{2}&m_{4}^{2}+m_{1}^{2}
\end{array}
\right|\nonumber\\
&\pm&\frac{1}{\lambda(s_{34},s_{12},q^{2})}\sqrt{4K(s_{34},s_{124};s_{12},m_{3}^2;m_4^2,q^2)K(s_{134},s_{12};s_{34},m_2^2;q^2,m_1^2)} \,,
\end{eqnarray}
Here, one has 
\begin{eqnarray}
K(x,y;z,u;v,w)&=&[x^2y\!+\!xy^2\!-\!xy(z\!+\!u\!+\!v\!+\!w)]\!+\![z^2u\!+\!zu^2\!-\!zu(x\!+\!y\!+\!v\!+\!w)]\nonumber\\
&+&[v^2w\!+\!vw^2\!-\!vw(x\!+\!y\!+\!z\!+\!u)]\!+\!x(zw\!+\!uv)\!+\!y(zv\!+\!uw)\,.\nonumber
\end{eqnarray}
\end{widetext}

\section{The ChPT form factors up to $O(p^4)$}
\label{appendix:M}
In this section, we present the form factors of the hadronization vector current calculated from $SU(3)$ ChPT up to NLO. 
In the following, the terms that are proportional to $q^\mu$ will be omitted, as they do not contribute to the production amplitude according to the \textit{Ward identity}. The form factors are given as 
\begin{widetext}
\begin{align}
&J^\mu_{(2)tree}(p_1,p_2,p_3,p_4)=\frac{s_{12}-m_\pi^2}{F_\pi^2}\biggl(\frac{2p_3^\mu}{2t_3-q^2}-\frac{2p_4^\mu}{2t_4-q^2}\biggr)\,, \\[3mm]
&J^\mu_{(4)tree}(p_1,p_2,p_3,p_4)
 =\frac{-1}{F_\pi^4}\biggl\{L_1^r\biggl[16(s_{12}-2m_\pi^2)^2\biggl(\frac{p_3^\mu}{t_1+t_2-t_3+t_4}-\frac{p_4^\mu}{t_1+t_2+t_3-t_4}\biggr)\biggr]\notag\\
&+L_2^r\biggl[4\biggl(2(t_3-t_4-2\nu)p_1^\mu+2(t_3-t_4+2\nu)p_2^\mu+\frac{(s_{12}^2+(-2t_1+t_3-t_4+2\nu)(2t_2-t_3+t_4+2\nu))p_3^\mu}{t_1+t_2-t_3+t_4}\notag\\
&-\frac{(s_{12}^2-(2t_1+t_3-t_4+2\nu)(2t_2+t_3-t_4-2\nu))p_4^\mu}{t_1+t_2+t_3-t_4}\biggr)\biggr]\notag\\
&+L_3^r\biggl[8(s_{12}-2m_\pi^2)^2\biggl(\frac{p_3^\mu}{t_1+t_2-t_3+t_4}-\frac{p_4^\mu}{t_1+t_2+t_3-t_4}\biggr)\biggr]\notag\\
&+L_4^r\biggl[\frac{16}{3}\biggl(\frac{(2m_k^2(-6s_{12}+7m_\pi^2+t_1+t_2+t_3-t_4)+m_\pi^2(-12s_{12}+19m_\pi^2+t_1+t_2+t_3-t_4))p_4^\mu}{t_1+t_2+t_3-t_4}\notag\\
&-\frac{(2m_k^2(-6s_{12}+7m_\pi^2+t_1+t_2-t_3+t_4)+m_\pi^2(-12s_{12}+19m_\pi^2+t_1+t_2-t_3+t_4))p_3^\mu}{t_1+t_2-t_3+t_4}\biggr)
\biggr]\notag\\
&+L_5^r\biggl[\frac{16}{3}m_\pi^2\biggl(\frac{(-9s_{12}+13m_\pi^2+t_1+t_2+t_3-t_4)p_4^\mu}{t_1+t_2+t_3-t_4}-\frac{(-9s_{12}+13m_\pi^2+t_1+t_2-t_3+t_4)p_3^\mu}{t_1+t_2-t_3+t_4}\biggr)\biggr]\notag\\
&+L_6^r\biggl[\frac{32}{3}m_\pi^2(2m_k^2+7m_\pi^2)\biggl(\frac{p_3^\mu}{t_1+t_2-t_3+t_4}-\frac{p_4^\mu}{t_1+t_2+t_3-t_4}\biggr)\biggr]\notag\\
&+L_8^r\biggl[\frac{128}{3}m_\pi^4\biggl(\frac{p_3^\mu}{t_1+t_2-t_3+t_4}-\frac{p_4^\mu}{t_1+t_2+t_3-t_4}\biggr)\biggr]+L_9^r\biggl[\frac{4q^2(s_{12}-m_\pi^2)p_3^\mu}{t_1+t_2-t_3+t_4}-\frac{4q^2(s_{12}-m_\pi^2)p_4^\mu}{t_1+t_2+t_3-t_4}\biggr]\notag\\
&+8(s_{12}-m_\pi^2)\biggl(\frac{p_3^\mu}{t_1+t_2-t_3+t_4}-\frac{p_4^\mu}{t_1+t_2+t_3-t_4}\biggr)\bigl[L_4^r(2m_k^2+m_\pi^2)+L_5^rm_k^2\bigr]\notag\\
&-2m_\pi^2\biggl(\frac{p_3^\mu}{t_1+t_2-t_3+t_4}-\frac{p_4^\mu}{t_1+t_2+t_3-t_4}\biggr)\bigl[L_4^r(16m_k^2+8m_\pi^2)+8L_5^rm_\pi^2\notag\\
&-L_6^r(32m_k^2+16m_\pi^2)-16L_8^rm_\pi^2\bigr]\biggr\} \,,\\[3mm]
&J^\mu_{(4)loop}(p_1,p_2,p_3,p_4)
 =\frac{1}{F_\pi^4}\biggl\{-\frac{1}{3}(p_3-p_4)_\nu G_1^{\mu\nu}(q,m_\pi^2,m_\pi^2)-\frac{1}{6}(p_3-p_4)_\nu G_1^{\mu\nu}(q,m_k^2,m_k^2)\notag\\
&-\frac{i}{3}(2p_1\cdot p_2-p_1\cdot p_3-p_1\cdot p_4-p_2\cdot p_3-p_2\cdot p_4+2p_3\cdot p_4)G_2^\mu(q,m_k^2,m_k^2)\notag\\
&-(p_3-p_4)^\mu(16p_1\cdot p_2+12m_\pi^2)G_3(p_1+p_2,m_\pi^2,m_\pi^2)-(p_3-p_4)^\mu(5p_1\cdot p_2+3m_\pi^2)G_3(p_1+p_2,m_k^2,m_k^2)\notag\\
&+(p_1-p_3)_\nu G_1^{\mu\nu}(p_1+p_3,m_\pi^2,m_\pi^2)+\frac{(p_1-p_3)_\nu}{2}G_1^{\mu\nu}(p_1+p_3,m_k^2,m_k^2)\notag\\
&+(p_2-p_3)_\nu G_1^{\mu\nu}(p_2+p_3,m_\pi^2,m_\pi^2)+\frac{(p_2-p_3)_\nu}{2}G_1^{\mu\nu}(p_2+p_3,m_k^2,m_k^2)\notag\\
&-(p_1-p_4)_\nu G_1^{\mu\nu}(p_1+p_4,m_\pi^2,m_\pi^2)-\frac{(p_1-p_4)_\nu}{2}G_1^{\mu\nu}(p_1+p_4,m_k^2,m_k^2)\notag\\
&-(p_2-p_4)_\nu G_1^{\mu\nu}(p_2+p_4,m_\pi^2,m_\pi^2)-\frac{(p_2-p_4)_\nu}{2}G_1^{\mu\nu}(p_2+p_4,m_k^2,m_k^2)\notag\\
&-2p_1\cdot p_3(p_2+p_4)^\mu G_3(p_1+p_3,m_\pi^2,m_\pi^2)-2p_2\cdot p_3(p_1+p_4)^\mu G_3(p_2+p_3,m_\pi^2,m_\pi^2)\notag\\
&+2p_1\cdot p_4(p_2+p_3)^\mu G_3(p_1+p_4,m_\pi^2,m_\pi^2)+2p_2\cdot p_4(p_1+p_3)^\mu G_3(p_2+p_4,m_\pi^2,m_\pi^2)\notag\\
&+\frac{2(q-2p_3)_\nu(3m_\pi^2-3s_{12}+q^2-2t_3)}{3(2t_3-q^2)}G_1^{\mu\nu}(q,m_\pi^2,m_\pi^2)-\frac{2(q-2p_4)_\nu(3m_\pi^2-3s_{12}+q^2-2t_4)}{3(2t_4-q^2)}G_1^{\mu\nu}(q,m_\pi^2,m_\pi^2)\notag\\
&+\frac{(q-2p_3)_\nu(3m_\pi^2-3s_{12}+q^2-2t_3)}{3(2t_3-q^2)}G_1^{\mu\nu}(q,m_k^2,m_k^2)-\frac{(q-2p_4)_\nu(3m_\pi^2-3s+q^2-2t_4)}{3(2t_4-q^2)}G_1^{\mu\nu}(q,m_k^2,m_k^2)\notag\\
&+i\frac{(t_1-m_\pi^2)(3m_\pi^2-3s_{12}+2t_1+4t_2-2t_3-2t_4)}{3(-t_1+t_2+t_3+t_4)}G_2^\mu(q,m_k^2,m_k^2)\notag\\
&+i\frac{(t_2-m_\pi^2)(3m_\pi^2-3s_{12}+4t_1+2t_2-2t_3-2t_4)}{3(t_1-t_2+t_3+t_4)}G_2^\mu(q,m_k^2,m_k^2)\notag\\
&+i\frac{(m_\pi^2-t_3)(3m_\pi^2-3s_{12}+q^2-2t_3)}{3(q^2-2t_3)}G_2^\mu(q,m_k^2,m_k^2)-i\frac{(m_\pi^2-t_4)(3m_\pi^2-3s_{12}+q^2-2t_4)}{3(q^2-2t_4)}G_2^\mu(q,m_k^2,m_k^2)\notag\\
&+\frac{2p_3^\mu(p_1-p_4)_\nu(q+p_2-p_3)_\rho}{2t_3-q^2}G_1^{\rho\nu}(p_1+p_4,m_\pi^2,m_\pi^2)+\frac{2p_3^\mu(p_2-p_4)_\nu(q+p_1-p_3)_\rho}{2t_3-q^2}G_1^{\rho\nu}(p_2+p_4,m_\pi^2,m_\pi^2)\notag\\
&-\frac{2p_4^\mu(p_1-p_3)_\nu(q+p_2-p_4)_\rho}{2t_4-q^2}G_1^{\rho\nu}(p_1+p_3,m_\pi^2,m_\pi^2)-\frac{2p_4^\mu(p_2-p_3)_\nu(q+p_1-p_4)_\rho}{2t_4-q^2}G_1^{\rho\nu}(p_2+p_3,m_\pi^2,m_\pi^2)\notag\\
&+\frac{p_3^\mu(p_1-p_4)_\nu(q+p_2-p_3)_\rho}{2t_3-q^2}G_1^{\rho\nu}(p_1+p_4,m_k^2,m_k^2)+\frac{p_3^\mu(p_2-p_4)_\nu(q+p_1-p_3)_\rho}{2t_3-q^2}G_1^{\rho\nu}(p_2+p_4,m_k^2,m_k^2)\notag\\
&-\frac{p_4^\mu(p_1-p_3)_\nu(q+p_2-p_4)_\rho}{2t_4-q^2}G_1^{\rho\nu}(p_1+p_3,m_k^2,m_k^2)-\frac{p_4^\mu(p_2-p_3)_\nu(q+p_1-p_4)_\rho}{2t_4-q^2}G_1^{\rho\nu}(p_2+p_3,m_k^2,m_k^2)\notag\\
&+\frac{2p_3^\mu(p_2\cdot p_3-t_2)}{2t_3-q^2}4p_1\cdot p_4G_3(p_1+p_4,m_\pi^2,m_\pi^2)+\frac{2p_3^\mu(p_1\cdot p_3-t_1)}{2t_3-q^2}4p_2\cdot p_4G_3(p_2+p_4,m_\pi^2,m_\pi^2)\notag\\
&-\frac{2p_4^\mu(p_2\cdot p_4-t_2)}{2t_4-q^2}4p_1\cdot p_3G_3(p_1+p_3,m_\pi^2,m_\pi^2)-\frac{2p_4^\mu(p_1\cdot p_4-t_1)}{2t_4-q^2}4p_2\cdot p_3G_3(p_2+p_3,m_\pi^2,m_\pi^2)\notag\\
&+\frac{2p_3^\mu}{2t_3-q^2}2[3m_\pi^4+4m_\pi^2(p_3\cdot p_4-t_4)+4p_1\cdot p_2(m_\pi^2-t_4+p_3\cdot p_4)]G_3(p_1+p_2,m_\pi^2,m_\pi^2)\notag\\
&-\frac{2p_4^\mu}{2t_4-q^2}2[3m_\pi^4+4m_\pi^2(p_3\cdot p_4-t_3)+4p_1\cdot p_2(m_\pi^2-t_3+p_3\cdot p_4)]G_3(p_1+p_2,m_\pi^2,m_\pi^2)\notag\\
&+\frac{2p_3^\mu}{2t_3-q^2}[2(p_3\cdot p_4-t_4+m_\pi^2)(p_1\cdot p_2+m_\pi^2)+2(p_3\cdot p_4-t_4)(p_1\cdot p_2)]G_3(p_1+p_2,m_k^2,m_k^2)\notag\\
&-\frac{2p_4^\mu}{2t_4-q^2}[2(p_3\cdot p_4-t_3+m_\pi^2)(p_1\cdot p_2+m_\pi^2)+2(p_3\cdot p_4-t_3)(p_1\cdot p_2)]G_3(p_1+p_2,m_k^2,m_k^2)\notag\\
&+\frac{4m_\pi^4p_3^\mu}{9(2t_3-q^2)}G_3(p_1+p_2,m_\eta^2,m_\eta^2)-\frac{4m_\pi^4p_4^\mu}{9(2t_4-q^2)}G_3(p_1+p_2,m_\eta^2,m_\eta^2)\notag\\
&-2(p_3-p_4)_\nu(m_\pi^2+2p_1\cdot p_2)G_5^{\mu\nu}(q,p_3+p_4,m_\pi^2,m_\pi^2,m_\pi^2)\notag\\
&-\frac{(p_3-p_4)_\nu}{2}(m_\pi^2+p_1\cdot p_2)G_5^{\mu\nu}(q,p_3+p_4,m_k^2,m_k^2,m_k^2)\notag\\
&-i[2p_1\cdot p_2 p_3\cdot p_4+m_\pi^2(p_1\cdot p_2+p_3\cdot p_4)]G_6^\mu(q,p_3+p_4,m_k^2,m_k^2,m_k^2)\notag\\
&-i\frac{3}{4}(p_1-p_3)_\nu(p_2-p_4)_\lambda G_4^{\mu\nu\lambda}(q,p_1+p_3,m_\pi^2,m_\pi^2,m_\pi^2)
 -i\frac{3}{4}(p_2-p_3)_\nu(p_1-p_4)_\lambda G_4^{\mu\nu\lambda}(q,p_2+p_3,m_\pi^2,m_\pi^2,m_\pi^2)\notag\\
&+i\frac{3}{4}(p_1-p_4)_\nu(p_2-p_3)_\lambda G_4^{\mu\nu\lambda}(q,p_1+p_4,m_\pi^2,m_\pi^2,m_\pi^2)
 +i\frac{3}{4}(p_2-p_4)_\nu(p_1-p_3)_\lambda G_4^{\mu\nu\lambda}(q,p_2+p_4,m_\pi^2,m_\pi^2,m_\pi^2)\notag\\
&-i\frac{3}{8}(p_1-p_3)_\nu(p_2-p_4)_\lambda G_4^{\mu\nu\lambda}(q,p_1+p_3,m_k^2,m_k^2,m_k^2)
 -i\frac{3}{8}(p_2-p_3)_\nu(p_1-p_4)_\lambda G_4^{\mu\nu\lambda}(q,p_2+p_3,m_k^2,m_k^2,m_k^2)\notag\\
&+i\frac{3}{8}(p_1-p_4)_\nu(p_2-p_3)_\lambda G_4^{\mu\nu\lambda}(q,p_1+p_4,m_k^2,m_k^2,m_k^2)
 +i\frac{3}{8}(p_2-p_4)_\nu(p_1-p_3)_\lambda G_4^{\mu\nu\lambda}(q,p_2+p_4,m_k^2,m_k^2,m_k^2)\notag\\
&-p_2\cdot p_4(p_1-p_3)_\nu G_5^{\mu\nu}(q,p_1+p_3,m_\pi^2,m_\pi^2,m_\pi^2)-p_1\cdot p_4(p_2-p_3)_\nu G_5^{\mu\nu}(q,p_2+p_3,m_\pi^2,m_\pi^2,m_\pi^2)\notag\\
&+p_2\cdot p_3(p_1-p_4)_\nu G_5^{\mu\nu}(q,p_1+p_4,m_\pi^2,m_\pi^2,m_\pi^2)+p_1\cdot p_3(p_2-p_4)_\nu G_5^{\mu\nu}(q,p_2+p_4,m_\pi^2,m_\pi^2,m_\pi^2)\notag\\
&-ip_1\cdot p_3 p_2\cdot p_4G_6^{\mu}(-p_1-p_3,p_2+p_4,m_\pi^2,m_\pi^2,m_\pi^2)-ip_2\cdot p_3 p_1\cdot p_4G_6^{\mu}(-p_2-p_3,p_1+p_4,m_\pi^2,m_\pi^2,m_\pi^2)\notag\\
&+ip_1\cdot p_4 p_2\cdot p_3G_6^{\mu}(-p_1-p_4,p_2+p_3,m_\pi^2,m_\pi^2,m_\pi^2)+ip_2\cdot p_4 p_1\cdot p_3G_6^{\mu}(-p_2-p_4,p_1+p_3,m_\pi^2,m_\pi^2,m_\pi^2)\notag\\
&+\frac{(t_3-t_4)\bigl(-m_\pi^2A^r(m_\eta^2)+(36s_{12}-33m_\pi^2)A^r(m_\pi^2)+18(s_{12}-m_\pi^2)A^r(m_k^2)\bigr)}{18(t_1+t_2+t_3-t_4)(t_1+t_2-t_3+t_4)}(p_1^\mu+p_2^\mu)\notag\\
&-\frac{p_3^\mu}{36(t_1+t_2+t_3-t_4)(t_1+t_2-t_3+t_4)}\biggl(2(t_3-t_4)m_\pi^2A^r(m_\eta^2)\notag\\
&+6A^r(m_\pi^2)\bigl(3(-t_1^2-2t_1t_2-t_2^2+t_3^2+2s_{12}(5t_1+5t_2+3t_3-3t_4)-2t_3t_4+t_4^2)-(36t_1+36t_2+25t_3-25t_4)m_\pi^2\bigr)\notag\\
&+9A^r(m_k^2)\bigl(-t_1^2-2t_1t_2-t_2^2+t_3^2+2s_{12}(5t_1+5t_2+3t_3-3t_4)-2t_3t_4+t_4^2-4(3t_1+3t_2+2t_3-2t_4)m_\pi^2\bigr)\biggr)\notag\\
&+\frac{p_4^\mu}{36(t_1+t_2+t_3-t_4)(t_1+t_2-t_3+t_4)}\biggl(2(-t_3+t_4)m_\pi^2A^r(m_\eta^2)\notag\\
&+6A^r(m_\pi^2)\bigl(3(-t_1^2-2t_1t_2-t_2^2+t_3^2+2s_{12}(5t_1+5t_2-3t_3+3t_4)-2t_3t_4+t_4^2)-(36t_1+36t_2-25t_3+25t_4)m_\pi^2\bigr)\notag\\
&+9A^r(m_k^2)\bigl(-t_1^2-2t_1t_2-t_2^2+t_3^2+2s_{12}(5t_1+5t_2-3t_3+3t_4)-2t_3t_4+t_4^2
 -4(3t_1+3t_2-2t_3+2t_4)m_\pi^2\bigr)\biggr)\biggr\}\,.\nonumber\\
\end{align}
Here, the one-loop functional functions $G$ are given as :
\begin{eqnarray}
G_{1}^{\mu\nu}(p,\!M_P^2,\!M_Q^2)&=&p^{\mu}p^{\nu}a(p^2,\!M_P^2,\!M_Q^2)\!+\!g^{\mu\nu}b(p^2,\!M_P^2,\!M_Q^2), \nonumber\\
G_{2}^{\mu}(p,\!M_P^2,\!M_Q^2)&=&ip^{\mu}\!\biggl(\!\frac{1}{2}B^r(p^2\!,\!M_P^2,\!M_Q^2)\!-\!B^r_{11}(p^2\!,\!M_P^2,\!M_Q^2)\!\biggr)\!, \nonumber\\
G_{3}(p,\!M_P^2,\!M_Q^2)&=&\frac{1}{4}B^r(p^2,\!M_P^2,\!M_Q^2),\nonumber\\
G^{\mu\nu\lambda}_{4}(p_a,p_b,\!M_P^2,\!M_Q^2,\!M_R^2)&=&i\bigl(
c_ap_{a\mu}p_{a\nu}p_{a\lambda}\!+\!c_bp_{b\mu}p_{b\nu}p_{b\lambda}\!+\!d_ap_{a\mu}p_{a\nu}p_{b\lambda} \!+\!d_bp_{b\mu}p_{b\nu}p_{a\lambda}\!+\!e_ap_{a\mu}p_{b\nu}p_{a\lambda}\nonumber\\
&+&\!e_bp_{a\mu}p_{b\nu}p_{b\lambda}\!+\!f_ap_{b\mu}p_{a\nu}p_{a\lambda}\!+\!f_bp_{b\mu}p_{a\nu}p_{b\lambda}\!+\!g_ap_{a\mu}g^{\nu\lambda}\!+\!g_bp_{b\nu}g^{\mu\lambda}\!+\!h_ap_{a\mu}g^{\nu\lambda}\nonumber\\
&+&\!h_bp_{b\mu}g^{\nu\lambda}\!+\!i_ap_{a\lambda}g^{\mu\nu}
\!+\!i_bp_{b\lambda}g^{\mu\nu}\bigr), \nonumber\\
G^{\mu\nu}_{5}(p_a,p_b,\!M_P^2,\!M_Q^2,\!M_R^2)&=&j_ag^{\mu\nu}\!+\!k_ap_{a\mu}p_{a\nu}\!+\!k_bp_{b\mu}p_{b\nu}+l_ap_{a\mu}p_{b\nu}\!+\!m_ap_{b\mu}p_{a\nu},\nonumber\\
G_{6}^\mu(p_a,p_b,\!M_P^2,\!M_Q^2,\!M_R^2)&=&i\bigl(n_ap_{a\mu}+n_bp_{b\mu}\bigr), \nonumber\\
G_{7}(p_a,p_b,\!M_P^2,\!M_Q^2,\!M_R^2)&=&\frac{1}{6}C(p_a^2,p_b^2,p_a\!\cdot\! p_b,\!M_P^2,\!M_Q^2,\!M_R^2),
\end{eqnarray}
where $P,Q,R$ represent different pseudoscalars. The abbreviated functions are given as 
\begin{align}
&a(p^2,\!M_P^2,\!M_Q^2)=\frac{1}{4}\biggl\{
4\bigl(\!B^r_{11}(p^2\!,\!M_P^2,M_Q^2)\!\!-\!\!B^r_{22}(p^2\!,\!M_P^2,M_Q^2))\!\!-\!\!B^r(p^2\!,\!M_P^2,M_Q^2)\!\biggr\},\notag\\
&b(p^2,\!M_P^2,\!M_Q^2)=-B^r_{20}(p^2,M_P^2,M_Q^2),\notag\\
&c_a=\frac{1}{3}\biggl\{C^r_{11}(p_a^2,p_b^2,p_a\cdot p_b,M_P^2,M_Q^2,M_R^2)-4C^r_{22}(p_a^2,p_b^2,p_a\cdot p_b,M_P^2,M_Q^2,M_R^2)+4C^r_{33}(p_a^2,p_b^2,p_a\cdot p_b,M_P^2,M_Q^2,M_R^2)\biggr\},\notag\\
&d_a=\frac{1}{3}\biggl\{C^r_{11}(p_a^2,p_b^2,p_a\cdot p_b,M_P^2,M_Q^2,M_R^2)-2C^r_{22}(p_a^2,p_b^2,p_a\cdot p_b,M_P^2,M_Q^2,M_R^2)-2\tilde{C}^r_{22}(p^2\!,\!M_P^2,M_Q^2)\notag\\
&\;\;\;\;+4\tilde{C}^r_{33}(p_a^2,p_b^2,p_a\cdot p_b,M_P^2,M_Q^2,M_R^2)\biggr\},\notag\\
&e_a=\frac{1}{3}\biggl\{-C/2+2C^r_{11}(p_a^2,p_b^2,p_a\cdot p_b,M_P^2,M_Q^2,M_R^2)+C^r_{11}(p_b^2,p_a^2,p_a\cdot p_b,M_P^2,M_R^2,M_Q^2)\notag\\
&\;\;\;\;-2C^r_{22}(p_a^2,p_b^2,p_a\cdot p_b,M_P^2,M_Q^2,M_R^2)-4\tilde{C}^r_{22}(p_a^2,p_b^2,p_a\cdot p_b,M_P^2,M_Q^2,M_R^2)+4\tilde{C}^r_{33}(p_a^2,p_b^2,p_a\cdot p_b,M_P^2,M_Q^2,M_R^2)\biggr\},\notag\\
&f_a=\frac{-2}{3}\biggl\{-2(\tilde{C}^r_{22}(p_a^2,p_b^2,p_a\cdot p_b,M_P^2,M_Q^2,M_R^2)-2\tilde{C}^r_{33}(p_a^2,p_b^2,p_a\cdot p_b,M_P^2,M_Q^2,M_R^2))\biggr\},\notag\\
&g_a=\frac{-2}{3}\biggl\{-2(C^r_{20}(p_a^2,p_b^2,p_a\cdot p_b,M_P^2,M_Q^2,M_R^2)-2C^r_{31}(p_a^2,p_b^2,p_a\cdot p_b,M_P^2,M_Q^2,M_R^2))\biggr\},\notag\\
&h_a=\frac{4}{3}\biggl\{C^r_{31}(p_a^2,p_b^2,p_a\cdot p_b,M_P^2,M_Q^2,M_R^2)\biggr\},\notag\\
&i_a=\frac{-2}{3}\biggl\{C^r_{20}(p_a^2,p_b^2,p_a\cdot p_b,M_P^2,M_Q^2,M_R^2)-2C^r_{31}(p_a^2,p_b^2,p_a\cdot p_b,M_P^2,M_Q^2,M_R^2)\biggr\},\notag\\
&j_a=-2C^r_{20}(p_a^2,p_b^2,p_a\cdot p_b,M_P^2,M_Q^2,M_R^2),\notag\\
&k_a=C^r_{11}(p_a^2,p_b^2,p_a\cdot p_b,M_P^2,M_Q^2,M_R^2)-2C^r_{22}(p_a^2,p_b^2,p_a\cdot p_b,M_P^2,M_Q^2,M_R^2),\notag\\
&l_a=-\frac{1}{2}C^r(p_a^2,p_b^2,p_a\cdot p_b,M_P^2,M_Q^2,M_R^2)+C^r_{11}(p_a^2,p_b^2,p_a\cdot p_b,M_P^2,M_Q^2,M_R^2)\notag\\
&\;\;\;+C^r_{11}(p_b^2,p_a^2,p_a\cdot p_b,M_P^2,M_R^2,M_Q^2)-2\tilde{C}^r_{22}r(p_a^2,p_b^2,p_a\cdot p_b,M_P^2,M_Q^2,M_R^2),\notag\\
&m_a=-2\tilde{C}^r_{22}(p_a^2,p_b^2,p_a\cdot p_b,M_P^2,M_Q^2,M_R^2),\notag\\
&n_a=\frac{1}{2}C^r(p_a^2,p_b^2,p_a\cdot p_b,M_P^2,M_Q^2,M_R^2)-C^r_{11}(p_a^2,p_b^2,p_a\cdot p_b,M_P^2,M_Q^2,M_R^2).
\end{align}
The above functions with subscripts \lq b' are similar to those with subscript \lq a', with only $p_a \leftrightarrow p_b$ and $M_Q \leftrightarrow M_R$. For example, one has 
\begin{equation}
 h_a=\frac{4}{3}C^r_{31}(p_a^2,p_b^2,p_a\cdot p_b,M_P^2,M_Q^2,M_R^2), \;\;\;\;
 h_b=\frac{4}{3}(p_b^2,p_a^2,p_a\cdot p_b,M_P^2,M_R^2,M_Q^2). 
\end{equation}
Indeed, all the finite parts of the loop functions are finally expressed by the $A$, $B$ and $C$ functions. The $A$ and $B$ functions can be written as
\begin{eqnarray}
 A^r(M_P^2)&=&\frac{M_p^2}{(4\pi)^2}\biggl(1-\ln\frac{M_P^2}{\mu^2}\biggr), \nonumber\\
 B^r(p^2,\!M_P^2,\!M_Q^2)&=&\frac{1}{16\pi^2}\!\biggl\{2\!-\!\ln\biggl(\frac{M_Q^2}{\mu^2}\biggr)
 +\!\frac{M_P^2\!-\!M_Q^2\!+\!p^2}{2p^2}\ln\biggl(\frac{M_Q^2}{M_P^2}\biggr) \nonumber\\
&+&\!\frac{\sqrt{\lambda(p^2\!,\!M_P^2,M_Q^2)}}{2p^2}
\times\ln\!\biggl(\!\frac{M_P^2\!+\!M_Q^2\!-\!p^2\!+\!\sqrt{\lambda(p^2,M_P^2,M_Q^2)}}{M_P^2\!+\!M_Q^2\!-\!p^2\!-\!\sqrt{\lambda(p^2,M_P^2,M_Q^2)}}\!\biggr)\!\biggr\},\nonumber\\
 B^r_{11}(p^2,\!M_P^2,\!M_Q^2)&=&\frac{-\!A^r(M_P^2)\!+\!A^r(M_Q^2)\!+\!B^r(p^2,\!M_P^2,\!M_Q^2)\bigl(M_P^2\!-\!M_Q^2\!+\!p^2\bigr)}{2p^2}, \nonumber\\
 B^r_{20}(p^2,\!M_P^2,\!M_Q^2)&=&-\frac{p^2-3M_P^2-3M_Q^2}{288\pi^2}
 +\!\frac{A^r(M_Q^2)\!+\!2B^r(p^2,\!M_P^2,\!M_Q^2)M_P^2\!-\!\Bigl(\!M_P^2\!-\!M_Q^2\!+\!p^2\!\Bigr)B^r_{11}(p^2,\!M_P^2,\!M_Q^2)}{6}, \nonumber\\
 B^r_{22}(p^2,\!M_P^2,\!M_Q^2)&=&-\frac{p^2-3M_P^2-3M_Q^2}{288\pi^2}
 +\!\frac{A^r(M_Q^2)\!\!-\!\!B^r(p^2,\!M_P^2,\!M_Q^2)M_P^2\!+\!2\Bigl(\!M_P^2\!\!-\!\!M_Q^2\!+\!p^2\!\Bigr)B^r_{11}(p^2,\!M_P^2,\!M_Q^2)}{6}. \nonumber\\
\end{eqnarray}
 with $\lambda(x,y,z)=(x+y-z)^2-4xy$. 
 The $C$ function without a subscript is defined as:
\begin{equation}
C(p_a^2,p_b^2,p_a\cdot p_b,\!M_P^2,\!M_Q^2,\!M_R^2)=-\frac{1}{16\pi^2}\frac{1}{\sqrt{\lambda}}\times\sum_{i=1}^{3}\sum_{\sigma=\pm}\biggl(Li_2\biggl(\frac{x_i}{x_i-z_i^\sigma}\biggr)-Li_2\biggl(\frac{x_i-1}{x_i-z_i^\sigma}\biggr)\biggr)\,,
\end{equation}
with $Li_2(z)=-\int_0^z\frac{dt}{t}\ln(1-t)$, and
\begin{eqnarray}
x_1&=&\frac{1}{2}\biggl(\!1\!+\!\frac{p_a^2\!-\!p_b^2\!-\!p_D^2}{\sqrt{\lambda(p_a^2,p_b^2,p_D^2)}}\biggr)\!+\!\frac{1}{2p_a^2}\biggl(\!-\!M_P^2\biggl(\!1\!+\!\frac{p_a^2\!+\!p_D^2\!-\!p_b^2}{\sqrt{\lambda(p_a^2,p_b^2,p_D^2)}}\!\biggr)\!+\!M_Q^2\biggl(\!1\!-\!\frac{p_a^2\!-\!p_D^2\!+\!p_b^2}{\sqrt{\lambda(p_a^2,p_b^2,p_D^2)}}\biggr)\!\biggr)+\frac{M_R^2}{\sqrt{\lambda(p_a^2,p_b^2,p_D^2)}},\nonumber\\
x_2&=&\frac{1}{2}\biggl(\!1\!+\!\frac{p_D^2\!-\!p_a^2\!-\!p_b^2}{\sqrt{\lambda(p_D^2,p_a^2,p_b^2)}}\biggr)\!+\!\frac{1}{2p_D^2}\biggl(\!-\!M_P^2\biggl(\!1\!+\!\frac{p_D^2\!+\!p_b^2\!-\!p_a^2}{\sqrt{\lambda(p_D^2,p_a^2,p_b^2)}}\!\biggr)
\!+\!M_Q^2\biggl(\!1\!-\!\frac{p_D^2\!-\!p_b^2\!+\!p_a^2}{\sqrt{\lambda(p_D^2,p_a^2,p_b^2)}}\biggr)\!\biggr)\!+\!\frac{M_R^2}{\sqrt{\lambda(p_D^2,p_a^2,p_b^2)}},\nonumber\\
x_3&=&\frac{1}{2}\biggl(\!1\!+\!\frac{p_b^2\!-\!p_D^2\!-\!p_a^2}{\sqrt{\lambda(p_b^2,p_D^2,p_a^2)}}\biggr)\!+\!\frac{1}{2p_b^2}\biggl(\!-\!M_P^2\biggl(\!1\!+\!\frac{p_b^2\!+\!p_a^2\!-\!p_D^2}{\sqrt{\lambda(p_b^2,p_D^2,p_a^2)}}\!\biggr)
\!+\!M_Q^2\biggl(\!1\!-\!\frac{p_b^2\!-\!p_a^2\!+\!p_D^2}{\sqrt{\lambda(p_b^2,p_D^2,p_a^2)}}\biggr)\!\biggr)\!+\!\frac{M_R^2}{\sqrt{\lambda(p_b^2,p_D^2,p_a^2)}},\nonumber\\
z_1^{\pm}&=&\frac{1}{2}\biggl(\!1\!-\!\frac{M_P^2\!-\!M_Q^2}{p_a^2}
\pm\sqrt{\bigl(\!1\!-\!\frac{M_P^2\!-\!M_Q^2}{p_a^2}\!\bigr)^2\!-\!\frac{4}{p_a^2}(M_Q^2\!-\!i\epsilon)}\biggr),\notag\\
z_2^{\pm}&=&\frac{1}{2}\biggl(\!1\!-\!\frac{M_P^2\!-\!M_Q^2}{p_D^2}
\pm\sqrt{\bigl(\!1\!-\!\frac{M_P^2\!-\!M_Q^2}{p_D^2}\!\bigr)^2\!-\!\frac{4}{p_D^2}(M_Q^2\!-\!i\epsilon)}\biggr),\notag\\
z_3^{\pm}&=&\frac{1}{2}\biggl(\!1\!-\!\frac{M_P^2\!-\!M_Q^2}{p_b^2}
\pm\sqrt{\bigl(\!1\!-\!\frac{M_P^2\!-\!M_Q^2}{p_b^2}\!\bigr)^2\!-\!\frac{4}{p_b^2}(M_Q^2\!-\!i\epsilon)}\biggr), \nonumber
\end{eqnarray}
where one has $p_D^2=p_a^2-2p_a\cdot p_b+p_b^2$ for simplicity. 
The definitions of other functions $C$ are given in Refs.~\cite{Passarino:1978jh,Unterdorfer:2005au},
\begin{align}
&C^r_{11}(p_a^2,p_b^2,p_a\!\cdot\!p_b,\!M_P^2,\!M_Q^2,\!M_R^2)=\frac{H_{11}(p_a^2,p_b^2,p_a\!\cdot\!p_b,\!M_P^2,\!M_Q^2,\!M_R^2)p_b^2\!-\!H_{11}(p_b^2,p_a^2,p_a\!\cdot\!p_b,\!M_P^2,\!M_R^2,\!M_Q^2)p_a\!\cdot\! p_b}{p_a^2p_b^2-p_a\cdot p_b},\notag\\
&C^r_{20}(p_a^2,p_b^2,p_a\!\cdot\!p_b,\!M_P^2,\!M_Q^2,\!M_R^2)=\!\frac{1}{64\pi^2}\!\!+
\frac{1}{4}\biggl\{2C(p_a^2,p_b^2,p_a\!\cdot\!p_b,\!M_P^2,\!M_Q^2,\!M_R^2)M_P^2\!\!+\!\!B^r(p_D^2,M_Q^2,M_R^2)\notag\\
&\;\;\;\;\;\;\;\;\;\;\;\;\;-\!\!(p_a^2\!+\!M_P^2\!-\!M_Q^2)C^r_{11a}(p_a^2,p_b^2,p_a\!\cdot\!p_b,\!M_P^2,\!M_Q^2,\!M_R^2)\!\!-\!\!(M_P^2\!\!-\!\!M_R^2\!+\!p_b^2)C^r_{11}(p_b^2,p_a^2,p_a\!\cdot\!p_b,\!M_P^2,\!M_R^2,\!M_Q^2)\biggr\},\notag\\
&C^r_{22}(p_a^2,p_b^2,p_a\!\cdot\!p_b,\!M_P^2,\!M_Q^2,\!M_R^2)=\!\frac{H_{21}(p_a^2,p_b^2,p_a\!\cdot\!p_b,\!M_P^2,\!M_Q^2,\!M_R^2)p_b^2\!-\!H_{22}(p_b^2,p_a^2,p_a\!\cdot\!p_b,\!M_P^2,\!M_R^2,\!M_Q^2)p_a\!\cdot\! p_b}{p_a^2p_b^2\!-\!p_a\cdot p_b},\notag\\
&\tilde{C}^r_{22}(p_a^2,p_b^2,p_a\!\cdot\!p_b,\!M_P^2,\!M_Q^2,\!M_R^2)=\!\frac{H_{22}(p_b^2,p_a^2,p_a\!\cdot\!p_b,\!M_P^2,\!M_R^2,\!M_Q^2)p_a^2\!-\!H_{21}(p_a^2,p_b^2,p_a\!\cdot\!p_b,\!M_P^2,\!M_Q^2,\!M_R^2)p_a\!\cdot\! p_b}{p_a^2p_b^2\!-\!p_a\cdot p_b},\notag\\
&C^r_{33}(p_a^2,p_b^2,p_a\!\cdot\!p_b,\!M_P^2,\!M_Q^2,\!M_R^2)=\!\frac{H_{31}(p_a^2,p_b^2,p_a\!\cdot\!p_b,\!M_P^2,\!M_Q^2,\!M_R^2)p_b^2\!-\!H_{32}(p_b^2,p_a^2,p_a\!\cdot\!p_b,\!M_P^2,\!M_R^2,\!M_Q^2)p_a\!\cdot\! p_b}{p_a^2p_b^2\!-\!p_a\cdot p_b},\notag\\
&\tilde{C}^r_{33}(p_a^2,p_b^2,p_a\!\cdot\!p_b,\!M_P^2,\!M_Q^2,\!M_R^2)=\!\frac{H_{32}(p_b^2,p_a^2,p_a\!\cdot\!p_b,\!M_P^2,\!M_R^2,\!M_Q^2)p_a^2\!-\!H_{31}(p_a^2,p_b^2,p_a\!\cdot\!p_b,\!M_P^2,\!M_Q^2,\!M_R^2)p_a\!\cdot\! p_b}{p_a^2p_b^2\!-\!p_a\cdot p_b},\notag\\
&C^r_{31}(p_a^2,p_b^2,p_a\!\cdot\!p_b,\!M_P^2,\!M_Q^2,\!M_R^2)=\!\frac{H_{30}(p_a^2,p_b^2,p_a\!\cdot\!p_b,\!M_P^2,\!M_Q^2,\!M_R^2)p_b^2\!-\!H_{30}(p_b^2,p_a^2,p_a\!\cdot\!p_b,\!M_P^2,\!M_R^2,\!M_Q^2)p_a\!\cdot\! p_b}{p_a^2p_b^2\!-\!p_a\cdot p_b},\notag\\
&H_{11}(p_a^2,p_b^2,p_a\!\cdot\!p_b,\!M_P^2,\!M_Q^2,\!M_R^2)
 =\!\frac{1}{2}\biggl\{\!-\!B^r(p_b^2,M_P^2,M_R^2)\!\!+\!\!B^r(p_D^2,M_Q^2,M_R^2)\notag\\
&\;\;\;\;\;\;\;\;\;\;\;\;\;+\!\!(p_a^2\!+\!M_P^2\!-\!M_Q^2)C(p_a^2,p_b^2,p_a\!\cdot\!p_b,\!M_P^2,\!M_Q^2,\!M_R^2)\biggr\},\notag\\
&H_{21}(p_a^2,p_b^2,p_a\!\cdot\!p_b,\!M_P^2,\!M_Q^2,\!M_R^2)
 =\frac{1}{2}\biggl[B^r(p_D^2,M_Q^2,M_R^2)+(p_a^2\!+\!M_P^2\!-\!M_Q^2)C^r_{11a}-\!B^r_{11}(p_D^2,M_Q^2,M_R^2)\biggr]\notag\\
&\;\;\;\;\;\;\;\;\;\;\;\;\;-C^r_{20}(p_a^2,p_b^2,p_a\!\cdot\!p_b,\!M_P^2,\!M_Q^2,\!M_R^2),\notag\\
&H_{22}(p_a^2,p_b^2,p_a\!\cdot\!p_b,\!M_P^2,\!M_Q^2,\!M_R^2)
 =\frac{1}{2}\biggl\{(p_a^2\!+\!M_P^2\!-\!M_Q^2)C^r_{11}(p_b^2,p_a^2,p_a\!\cdot\!p_b,\!M_P^2,\!M_R^2,\!M_Q^2)\!-\!B^r_{11}(p_b^2,M_P^2,M_R^2)\notag\\
&\;\;\;\;\;\;\;\;\;\;\;\;\;\!+\!B^r_{11}(p_D^2,M_Q^2,M_R^2)\biggr\},\notag\\
&H_{30}(p_a^2,p_b^2,p_a\!\cdot\!p_b,\!M_P^2,\!M_Q^2,\!M_R^2)
 =\frac{1}{2}\biggl\{ (p_a^2\!+\!M_P^2\!-\!M_Q^2)C^r_{20}(p_a^2,p_b^2,p_a\!\cdot\!p_b,\!M_P^2,\!M_Q^2,\!M_R^2)\!-\!B^r_{20}(p_b^2,M_P^2,M_R^2)\!\notag\\
&\;\;\;\;\;\;\;\;\;\;\;\;\;+\!B^r_{20}(p_D^2,M_Q^2,M_R^2)\biggr\},\notag\\
&H_{31}(p_a^2,p_b^2,p_a\!\cdot\!p_b,\!M_P^2,\!M_Q^2,\!M_R^2)
 =-\!2C^r_{31}(p_a^2,p_b^2,p_a\!\cdot\!p_b,\!M_P^2,\!M_Q^2,\!M_R^2)\notag\\
&\;\;\;\;\;\;\;\;\;\;\;\;\;+\!\frac{1}{2}\biggl[B^r(p_D^2,M_Q^2,M_R^2)\!+\!(p_a^2\!+\!M_P^2\!-\!M_Q^2)C^r_{22}\!\!-\!\!2B^r_{11}(p_D^2,M_Q^2,M_R^2)\!+\!B^r_{22}(p_D^2,M_Q^2,M_R^2)\biggr],\notag\\
&H_{32}(p_a^2,p_b^2,p_a\!\cdot\!p_b,\!M_P^2,\!M_Q^2,\!M_R^2)
 =\frac{1}{2}\biggl\{(p_a^2\!+\!M_P^2\!-\!M_Q^2)C^r_{22}(p_b^2,p_a^2,p_a\!\cdot\!p_b,\!M_P^2,\!M_R^2,\!M_Q^2)\!-\!B^r_{22}(p_b^2,M_P^2,M_R^2)\!\notag\\
&\;\;\;\;\;\;\;\;\;\;\;\;\;+\!B^r_{22}(p_D^2,M_Q^2,M_R^2)\biggr\}.
\end{align}
\end{widetext}
Notice that we used the $\overline{MS}$ renormalization scheme, while the earlier studies \cite{Gasser:1983yg,Gasser:1984gg,Bijnens:2014lea} adopted the $\overline{MS}-1$ scheme. Therefore, one needs the following relation to transform the LECs, 
\begin{equation}
 L_i^{\overline{\rm MS}}=L_i^{\overline{\rm MS}-1}-\frac{\Gamma_i}{32\pi^2}\,, \nonumber
\end{equation}
where $\Gamma_i$ can be found in Ref.~\cite{Gasser:1984gg}.

For the readers' convenience, we briefly introduce how to deal with the loop integrals. Integrals with a tensor structure can be expressed as linear combinations of Lorentz-covariant tensors, which are built from the metric tensor $g_{\mu\nu}$ and a linearly independent set of momenta through the following functions \cite{Kniehl:1993ay,tHooft:1978jhc,Passarino:1978jh}. 
 The loop integral with one propagator is known as the $A$ function, 
\begin{equation}
 A(M_P^2)=\frac{1}{i}\mu^{4-D}\!\!\int\frac{d^Dk}{(2\pi)^D}\frac{1}{(k^2-M_P^2)}\,. 
\end{equation}
Note that the propagator is of the Feynman prescription. 
The loop integrals with two or three propagators are defined as
\begin{eqnarray}
\{X\}&\doteq&\!\frac{1}{i}\mu^{4-D}\!\!\int\frac{d^Dk}{(2\pi)^D}\frac{X}{(k^2-M_P^2)((k-p)^2-M_Q^2)}, \nonumber\\
\{\{X\}\}&\doteq&\!\frac{\mu^{4-D}}{i}\!\!\int\!\!\frac{d^Dk}{(2\pi)^D} \nonumber\\
&&\frac{X}{(k^2\!-\!M_P^2)((k\!-\!p_a)^2\!-\!M_Q^2)(\!(k\!-\!p_b)^2\!-\!M_R^2)}. \nonumber\\
\end{eqnarray}
One has 
\begin{eqnarray}
\{1\}&=&B(p^2), \nonumber\\
\{k_\mu\}&=&p_\mu B_{11}(p^2,M_P^2,M_Q^2), \nonumber\\
\{k_\mu k_\nu\}\!&=&g_{\mu\nu}B_{20}(p^2\!,\!M_P^2,M_Q^2)\!+\!p_\mu p_\nu B_{22}(p^2\!,\!M_P^2,M_Q^2), \nonumber\\
\{\{1\}\}&=&C(p_a^2,p_b^2,p_a\!\cdot\! p_b,M_P^2,M_Q^2,M_R^2), \nonumber\\
\{\{k_\mu\}\}&=&p_{a\mu}C_{11a}+p_{b\mu}C_{11b},\nonumber\\
\{\{k_\mu k_\nu\}\}&=&g_{\mu\nu}C_{20}+p_{a\mu}p_{a\nu}C_{22a}+p_{b\mu}p_{b\nu}C_{22b}\nonumber\\
&+&\!(p_{a\mu}p_{b\nu}+p_{b\mu}p_{a\nu})\tilde{C}_{22}, \nonumber\\
\{\{k_{\mu}k_{\nu}k_{\lambda}\}\}&=&p_{a\mu}p_{a\nu}p_{a\lambda}C_{33a}+p_{b\mu}p_{b\nu}p_{b\lambda}C_{33b}\nonumber\\
&+&\!(p_{a\mu}p_{a\nu}p_{b\lambda}\!+\!p_{a\mu}p_{b\nu}p_{a\lambda}\!+\!p_{b\mu}p_{a\nu}p_{a\lambda})\tilde{C}_{33a} \nonumber\\
&+&\!(p_{b\mu}p_{b\nu}p_{a\lambda}\!+\!p_{b\mu}p_{a\nu}p_{b\lambda}\!+\!p_{a\mu}p_{b\nu}p_{b\lambda})\tilde{C}_{33b}\nonumber\\
&+&\!(p_{a\mu}g_{\nu\lambda}+p_{a\nu}g_{\mu\lambda}+p_{a\lambda}g_{\mu\nu})C_{31a}\nonumber\\
&+&\!(p_{b\mu}g_{\nu\lambda}+p_{b\nu}g_{\mu\lambda}+p_{b\lambda}g_{\mu\nu})C_{31b},
\end{eqnarray}
with 
\begin{eqnarray}
 C_{ija}&=&C_{ij}(p_a^2,p_b^2,p_a\!\cdot\! p_b,M_P^2,M_Q^2,M_R^2), \nonumber\\ 
 C_{ijb}&=&C_{ij}(p_b^2,p_a^2,p_a\cdot p_b,M_P^2,M_Q^2,M_R^2). \nonumber
\end{eqnarray}
The finite parts of these functions after renormalization are given in the previous section. 

\section{The form factors from RChT} \label{appendix:RChT}
As discussed before, the form factors for the hadronization of the vector current can be separated into two parts: one is from the lowest order $\mathcal{O}(p^2)$ ChPT effective Lagrangians, and the other is from those with resonances. The form factors for $\gamma^*\to 4\pi$ with resonances can be written as 
\begin{widetext}
\begin{align}
 J^\mu_R=&\frac{2}{F_\pi^4}\biggl\{\frac{4p_3^\mu(c_dp_1\cdot p_2+2c_mm_\pi^2)^2}{3(2t_3-q^2)D_\sigma(s_{12})}-\frac{4p_4^\mu(c_dp_1\cdot p_2+2c_mm_\pi^2)^2}{3(2t_4-q^2)D_\sigma(s_{12})}\biggr\}\notag\\
&+\frac{1}{F_\pi^4}\biggl\{\frac{F_VG_Vq^2p_3^\mu(s_{12}-m_\pi^2)}{D_\rho(q^2)(2t_3-q^2)}+F_VG_V(t_2p_3^\mu-t_3p_2^\mu)\biggl(\frac{1}{D_\rho(q^2)}-\frac{1}{D_\rho(p_2\cdot p_3+2m_\pi^2)}\biggr)\notag\\
&+4G_V^2\biggl(\frac{2p_3^\mu(p_2\cdot p_4(p_1\cdot p_3-t_1)-\frac{1}{2}(s_{12}-2m_\pi^2)(p_3\cdot p_4-t_4))}{(2t_3-q^2)D_\rho(p_1\cdot p_4+2m_\pi^2)}+\frac{2p_3^\mu(\frac{1}{2}(s_{12}-2m_\pi^2)p_3^\mu-p_2^\mu p_1\cdot p_3)}{D_\rho(p_2\cdot p_3+2m_\pi^2)}\biggr)\biggr\}\notag\\
&+\frac{1}{F_\pi^4}\biggl\{\frac{F_VG_Vq^2p_3^\mu(s_{12}-m_\pi^2)}{D_\rho(q^2)(2t_3-q^2)}+F_VG_V(t_1p_3^\mu-t_3p_1^\mu)\biggl(\frac{1}{D_\rho(q^2)}-\frac{1}{D_\rho(p_1\cdot p_3+2m_\pi^2)}\biggr)\notag\\
&+4G_V^2\biggl(\frac{2p_3^\mu(p_1\cdot p_4(p_2\cdot p_3-t_2)-\frac{1}{2}(s_{12}-2m_\pi^2)(p_3\cdot p_4-t_4))}{(2t_3-q^2)D_\rho(p_2\cdot p_4+2m_\pi^2)}+\frac{2p_3^\mu(\frac{1}{2}(s_{12}-2m_\pi^2)p_3^\mu-p_1^\mu p_2\cdot p_3)}{D_\rho(p_1\cdot p_3+2m_\pi^2)}\biggr)\biggr\}\notag\\
&-\frac{1}{F_\pi^4}\biggl\{\frac{F_VG_Vq^2p_4^\mu(s_{12}-m_\pi^2)}{D_\rho(q^2)(2t_4-q^2)}+F_VG_V(t_1p_4^\mu-t_4p_1^\mu)\biggl(\frac{1}{D_\rho(q^2)}-\frac{1}{D_\rho(p_1\cdot p_4+2m_\pi^2)}\biggr)\notag\\
&+4G_V^2\biggl(\frac{2p_4^\mu(p_1\cdot p_3(p_2\cdot p_4-t_2)-\frac{1}{2}(s_{12}-2m_\pi^2)(p_3\cdot p_4-t_3))}{(2t_4-q^2)D_\rho(p_2\cdot p_3+2m_\pi^2)}+\frac{2p_3^\mu(\frac{1}{2}(s_{12}-2m_\pi^2)p_4^\mu-p_1^\mu p_2\cdot p_4)}{D_\rho(p_1\cdot p_4+2m_\pi^2)}\biggr)\biggr\}\notag\\
&-\frac{1}{F_\pi^4}\biggl\{\frac{F_VG_Vq^2p_4^\mu(s_{12}-m_\pi^2)}{D_\rho(q^2)(2t_4-q^2)}+F_VG_V(t_2p_4^\mu-t_4p_2^\mu)\biggl(\frac{1}{D_\rho(q^2)}-\frac{1}{D_\rho(p_2\cdot p_4+2m_\pi^2)}\biggr)\notag\\
&+4G_V^2\biggl(\frac{2p_4^\mu(p_2\cdot p_3(p_1\cdot p_4-t_1)-\frac{1}{2}(s_{12}-2m_\pi^2)(p_3\cdot p_4-t_3))}{(2t_4-q^2)D_\rho(p_1\cdot p_3+2m_\pi^2)}+\frac{2p_4^\mu(\frac{1}{2}(s_{12}-2m_\pi^2)p_4^\mu-p_2^\mu p_1\cdot p_4)}{D_\rho(p_2\cdot p_4+2m_\pi^2)}\biggr)\biggr\}\notag\\
&+\frac{1}{F_\pi^4}\biggl\{\frac{8F_VG_V^3q^2p_3^\mu\bigl((p_1\cdot p_3-t_1)p_2\cdot p_4-\frac{1}{2}(s_{12}-2m_\pi^2)(p_3\cdot p_4-t_4)\bigr)}{F_\pi^2(2t_3-q^2)D_\rho(q^2)D_\rho(p_1\cdot p_4+2m_\pi^2)}\notag\\
&\!+\!\frac{4G_V^2\bigl(p_3^\mu(m_\pi^2+p_1\cdot p_4)(p_2\!\cdot\!(p_1-p_4))+p_1^\mu(m_\pi^2+p_2\cdot p_3)((p_2-p_3)\cdot p_4)\bigr)}{D_\rho(p_2\cdot p_3+2m_\pi^2)D_\rho(p_1\cdot p_4+2m_\pi^2)}\notag\\
&\!+\!\frac{F_VG_V\bigl(p_4^\mu(p_1\!\cdot\!(p_3\!-\!p_2))q^2\!+\!p_1^\mu((p_2\!-\!p_3)\!\cdot\! p_4)q^2\!+\!((t_4\!-\!t_1)(p_2\!-\!p_3)^\mu\!-\!(t_3\!-\!t_2)(p_1\!-\!p_4)^\mu)(m_\pi^2\!+\!p_1\cdot p_4)\bigr)}{D_\rho(q^2)D_\rho(p_1\cdot p_4+2m_\pi^2)}\biggr\}\notag\\
&+\frac{1}{F_\pi^4}\biggl\{\frac{8F_VG_V^3q^2p_3^\mu\bigl((p_2\cdot p_3-t_2)p_1\cdot p_4-\frac{1}{2}(s_{12}-2m_\pi^2)(p_3\cdot p_4-t_4)\bigr)}{F_\pi^2(2t_3-q^2)D_\rho(q^2)D_\rho(p_2\cdot p_4+2m_\pi^2)}\notag\\
&\!+\!\frac{4G_V^2\bigl(p_3^\mu(m_\pi^2+p_2\cdot p_4)(p_1\!\cdot\!(p_2-p_4))+p_2^\mu(m_\pi^2+p_1\cdot p_3)((p_1-p_3)\cdot p_4)\bigr)}{D_\rho(p_1\cdot p_3+2m_\pi^2)D_\rho(p_2\cdot p_4+2m_\pi^2)}\notag\\
&\!+\!\frac{F_VG_V\bigl(p_4^\mu(p_2\!\cdot\!(p_3\!-\!p_1))q^2\!+\!p_2^\mu((p_1\!-\!p_3)\!\cdot\! p_4)q^2\!+\!((t_4\!-\!t_2)(p_1\!-\!p_3)^\mu\!-\!(t_3\!-\!t_1)(p_2\!-\!p_4)^\mu)(m_\pi^2\!+\!p_2\cdot p_4)\bigr)}{D_\rho(q^2)D_\rho(p_2\cdot p_4+2m_\pi^2)}\biggr\}\notag\\
&-\frac{1}{F_\pi^4}\biggl\{\frac{8F_VG_V^3q^2p_4^\mu\bigl((p_1\cdot p_4-t_1)p_2\cdot p_3-\frac{1}{2}(s_{12}-2m_\pi^2)(p_3\cdot p_4-t_3)\bigr)}{F_\pi^2(2t_4-q^2)D_\rho(q^2)D_\rho(p_1\cdot p_3+2m_\pi^2)}\notag\\
&\!+\!\frac{4G_V^2\bigl(p_4^\mu(m_\pi^2+p_1\cdot p_3)(p_2\!\cdot\!(p_1-p_3))+p_1^\mu(m_\pi^2+p_2\cdot p_4)((p_2-p_4)\cdot p_3)\bigr)}{D_\rho(p_2\cdot p_4+2m_\pi^2)D_\rho(p_1\cdot p_3+2m_\pi^2)}\notag\\
&\!+\!\frac{F_VG_V\bigl(p_3^\mu(p_1\!\cdot\!(p_4\!-\!p_2))q^2\!+\!p_1^\mu((p_2\!-\!p_4)\!\cdot\! p_3)q^2\!+\!((t_3\!-\!t_1)(p_2\!-\!p_4)^\mu\!-\!(t_4\!-\!t_2)(p_1\!-\!p_3)^\mu)(m_\pi^2\!+\!p_1\cdot p_3)\bigr)}{D_\rho(q^2)D_\rho(p_1\cdot p_3+2m_\pi^2)}\biggr\}\notag\\
&-\frac{1}{F_\pi^4}\biggl\{\frac{8F_VG_V^3q^2p_4^\mu\bigl((p_2\cdot p_4-t_2)p_1\cdot p_3-\frac{1}{2}(s_{12}-2m_\pi^2)(p_3\cdot p_4-t_3)\bigr)}{F_\pi^2(2t_4-q^2)D_\rho(q^2)D_\rho(p_2\cdot p_3+2m_\pi^2)}\notag\\
&\!+\!\frac{4G_V^2\bigl(p_4^\mu(m_\pi^2+p_2\cdot p_3)(p_1\!\cdot\!(p_2-p_3))+p_2^\mu(m_\pi^2+p_1\cdot p_4)((p_1-p_4)\cdot p_3)\bigr)}{D_\rho(p_1\cdot p_4+2m_\pi^2)D_\rho(p_2\cdot p_3+2m_\pi^2)}\notag\\
&\!+\!\frac{F_VG_V\bigl(p_3^\mu(p_2\!\cdot\!(p_4\!-\!p_1))q^2\!+\!p_2^\mu((p_1\!-\!p_4)\!\cdot\! p_3)q^2\!+\!((t_3\!-\!t_2)(p_1\!-\!p_4)^\mu\!-\!(t_4\!-\!t_1)(p_2\!-\!p_3)^\mu)(m_\pi^2\!+\!p_2\cdot p_3)\bigr)}{D_\rho(q^2)D_\rho(p_2\cdot p_3+2m_\pi^2)}\biggr\}\notag\\
&-\frac{16\sqrt{2}G_V\lambda_3^{SV}(p_4^\mu t_3-p_3^\mu t_4)(c_d p_1\cdot p_2+c_mm_\pi^2)}{F_\pi^4D_\rho(2p_3\cdot p_4+2m_\pi^2)D_\sigma(2p_1\cdot p_2+2m_\pi^2)}\notag\\
&+\frac{16\lambda^{SVV}G_V(p_4^\mu t_3-p_3^\mu t_4)(2\sqrt{2}(4\lambda_6^Vm_\pi^2-\lambda_{22}^Vq^2+F_V)(c_dp_1\cdot p_2+c_mm_\pi^2))}{F_\pi^4D_\rho(q^2)D_\rho(2p_3\cdot p_4+2m_\pi^2)D_\sigma(2p_1\cdot p_2+2m_\pi^2)}\notag\\
&-\frac{16\lambda_6^{VV}G_Vm_\pi^2(p_4^\mu t_3-p_3^\mu t_4)(2\sqrt{2}(4\lambda_6^Vm_\pi^2-\lambda_{22}^Vq^2))}{F_\pi^4D_\rho(q^2)D_\rho(2p_3\cdot p_4+2m_\pi^2)}\notag\\
&\!-\!\frac{2\sqrt{2}G_V(4\lambda_6^Vm_\pi^2\!-\!\lambda_{22}^Vq^2)}{F_\pi^4D_\rho(q^2)D_\rho(2p_1\cdot p_3+2m_\pi^2)}\biggl\{q^2(p_3^\mu(p_1\cdot p_4\!-\!p_1\cdot p_2)\!+\!p_1^\mu(p_2\cdot p_3\!-\!p_3\cdot p_4))\notag\\
&\!+\!(p_1\cdot p_3\!+\!m_\pi^2)((t_2\!-\!t_4)(p_1^\mu\!-\!p_3^\mu)\!-\!(t_1\!-\!t_3)(p_2^\mu\!-\!p_4^\mu))\biggr\}\notag\\
&+\frac{2\sqrt{2}G_V(4\lambda_6^Vm_\pi^2\!-\!\lambda_{22}^Vq^2)}{F_\pi^4D_\rho(q^2)D_\rho(2p_2\cdot p_3+2m_\pi^2)}\biggl\{q^2(p_3^\mu(p_1\cdot p_2\!-\!p_2\cdot p_4)\!+\!p_2^\mu(p_3\cdot p_4\!-\!p_1\cdot p_3))\notag\\
&\!+\!(p_2\cdot p_3\!+\!m_\pi^2)((t_2\!-\!t_3)(p_1^\mu\!-\!p_4^\mu)\!-\!(t_1\!-\!t_4)(p_2^\mu\!-\!p_3^\mu))\biggr\}\notag\\
&+\frac{2\sqrt{2}G_V(4\lambda_6^Vm_\pi^2\!-\!\lambda_{22}^Vq^2)}{F_\pi^4D_\rho(q^2)D_\rho(2p_1\cdot p_4+2m_\pi^2)}\biggl\{q^2(p_4^\mu(p_1\cdot p_3\!-\!p_1\cdot p_2)\!+\!p_1^\mu(p_2\cdot p_4\!-\!p_3\cdot p_4))\notag\\
&\!+\!(p_1\cdot p_4\!+\!m_\pi^2)((t_2\!-\!t_3)(p_1^\mu\!-\!p_4^\mu)\!-\!(t_1\!-\!t_4)(p_2^\mu\!-\!p_3^\mu))\biggr\}\notag\\
&-\frac{2\sqrt{2}G_V(4\lambda_6^Vm_\pi^2\!-\!\lambda_{22}^Vq^2)}{F_\pi^4D_\rho(q^2)D_\rho(2p_2\cdot p_4+2m_\pi^2)}\biggl\{q^2(p_4^\mu(p_1\cdot p_2\!-\!p_2\cdot p_3)\!+\!p_2^\mu(p_3\cdot p_4\!-\!p_1\cdot p_4))\notag\\
&\!+\!(p_2\cdot p_4\!+\!m_\pi^2)((t_2\!-\!t_4)(p_1^\mu\!-\!p_3^\mu)\!-\!(t_1\!-\!t_3)(p_2^\mu\!-\!p_4^\mu))\biggr\}\notag\\
&\!+\!\frac{16G_Vq^2(2\sqrt{2}(4\lambda_6^Vm_\pi^2-\lambda_{22}^Vq^2)+F_V)(c_dp_1\cdot p_2+c_mm_\pi^2)}{F_\pi^6D_\rho(q^2)D_\sigma(S_{12})}\biggl\{\frac{p_3^\mu(c_d(p_3\!\cdot\! p_4\!-\!t_4)+c_mm_\pi^2)}{q^2-2t_3}\notag\\
&\!-\!\frac{p_4^\mu(c_d(p_3\!\cdot\! p_4\!-\!t_3)+c_mm_\pi^2)}{q^2-2t_4}\biggr\}\,.
\end{align}
\end{widetext}
The denominators of the propagators with energy-dependent widths are given as 
\begin{equation}
 D_P(t)=M_P^2-t-iM_P\Gamma_P(t), \nonumber
\end{equation}
and the widths are 
\begin{eqnarray}
\Gamma_\rho(t)\!&=&\!\frac{M_\rho t}{96\pi F_\pi^2}\!\biggl[\rho_\pi^{3}\!(t)\theta(t\!-\!4m_\pi^2)\!+\!\frac{1}{2}\rho_K^{3}(t)\theta(t\!-\!4m_k^2)\biggr], \nonumber\\
\Gamma_S(t)\!&=&\!\Gamma_S\frac{t\rho_\pi(t)}{M_S^2\rho_\pi(M_S^2)}\theta(t-4M_\pi^2).\nonumber
\end{eqnarray}
Here, $\rho_P(t)=\sqrt{1-4m_P^2/t}$ is the two-body phase space factor. 
$S$ represent the lightest scalars, $\sigma$, $f_0(980)$. The width of the $f_0(980)$ are taken from Refs.\cite{Dai:2014lza,Dai:2014zta}, and others are taken from PDG \cite{ParticleDataGroup:2022pth}. Note that we take the Breit-Wigner mass and width for $\sigma$  \cite{ParticleDataGroup:2022pth}.
The decay widths involving the lightest vector resonances are given as 
\begin{widetext}
\begin{eqnarray}
\Gamma_{\rho\rightarrow e^+e^-}&=&\frac{4\pi\alpha^2}{3M_\rho}\biggl(F_V+2\sqrt{2}(\lambda_6^Vm_\pi^2-\lambda_{22}^VM_\rho^2)^2\biggr)\times\biggl(1+\frac{2m_e^2}{M_\rho^2}\biggr)\biggl(1-\frac{4m_e^2}{M_\rho^2}\biggr)^{\frac{1}{2}}\!,\nonumber\\
\Gamma_{\omega\rightarrow e^+e^-}&=&\!\frac{4\pi\alpha^2}{81M_\omega}\!\biggl[\!\sin\!\theta_V\!\biggl(\!2\sqrt{2}[4\lambda_6^V(m_\pi^2\!-\!4m_k^2)\!+\!3\lambda_{22}^V M_\omega^2]\!-\!3F_V\!\biggr)\!+\!32\lambda_6^V\!\cos\!\theta_V\!(m_k^2\!-\!m_\pi^2)\biggr]^2 \!\biggl(\!1+\!\frac{2m_e^2}{M_\omega^2}\!\biggr)\biggl(\!1\!-\!\frac{4m_e^2}{M_\omega^2}\!\biggr)^{\frac{1}{2}}\!,\nonumber\\
\Gamma_{\phi\rightarrow e^+e^-}&=&\!\frac{4\pi\alpha^2}{81M_\phi}\!\biggl[\!\cos\!\theta_V\!\biggl(\!3F_V\!-\!2\sqrt{2}[3\lambda_{22}^VM_\phi^2\!+\!4\lambda_6^V(m_\pi^2\!-\!4m_k^2)]\!\biggr)\!+\!32\lambda_6^V\!\sin\!\theta_V\!(m_k^2\!-\!m_\pi^2)\biggr]^2\biggl(\!1\!+\!\frac{2m_e^2}{M_\phi^2}\!\biggr)\biggl(\!1\!-\!\frac{4m_e^2}{M_\phi^2}\!\biggr)^{\frac{1}{2}}\!, \nonumber\\
\Gamma_{\phi\rightarrow a_0^0\gamma}&=&\frac{\alpha(M_{a_0^0}^2-M_\phi^2)^3}{18M_\phi^5D_\rho(0)^2}\biggl\{\lambda_3^{SV}F_V(\sqrt{2}\cos\theta_V-2\sin\theta_V)^2\!+\!2(\cos\theta_V\!-\!\sqrt{2}\sin\theta_V)(8\lambda_6^V\lambda^{SVV}m_\pi^2\!-\!D_\rho(0)\lambda_3^{SV})\biggr\}^2
\!,\nonumber\\
\Gamma_{\phi\rightarrow f_0\gamma}&=&\frac{\alpha(M_{f_0}^2-M_\phi^2)^3}{5832M_\phi^5D_\omega(0)^2D_\phi(0)^2}\biggl\{2\lambda^{SVV}(D_\omega(0)\!-\!D_\phi(0))\sin(3\theta_V)\!\times\!(3\sqrt{3}F_V+80\lambda_6^Vm_k^2-32\lambda_6^Vm_\pi^2) \nonumber\\
&+&\lambda^{SVV}\!(D_\omega(0)\!-\!D_\phi(0))\cos(3\theta_V)\biggl(\!3F_V\!+\!8\sqrt{2}\lambda_6^V\!(7m_\pi^2\!-\!4m_k^2)\!\biggr)\!+\!6\sin(\theta_V)\biggl(\!\sqrt{2}\lambda^{SVV}\!F_V(D_\omega(0)\!+\!D_\phi(0)) \nonumber\\
&+&16\lambda_6^V\lambda^{SVV}m_k^2(3D_\omega(0)+D_\phi(0))-4D_\omega(0)(D_\phi(0)\lambda_3^{SV}+8\lambda_6^V\lambda^{SVV}m_\pi^2)\biggr) \nonumber\\
&+&3\cos(\theta_V)\biggl(\lambda^{SVV}F_V(7D_\omega(0)+D_\phi(0))+8\sqrt{2}\bigl(4\lambda_6^V\lambda^{SVV}m_k^2(3D_\omega(0)+D_\phi(0)) \nonumber\\
&-&\lambda_6^V\lambda^{SVV}m_\pi^2(5D_\omega(0)+3D_\phi(0))-D_\omega(0)D_\phi(0)\lambda_3^{SV}\bigr)\biggr)\biggr\}^2\,. 
\end{eqnarray}
\end{widetext}

\section{The amplitudes from EMC}
We also test the contributions from the higher-order electromagnetic corrections (EMCs).
\label{appendixF}
\begin{figure}[!htb]
\centering
\includegraphics[width=0.48\textwidth]{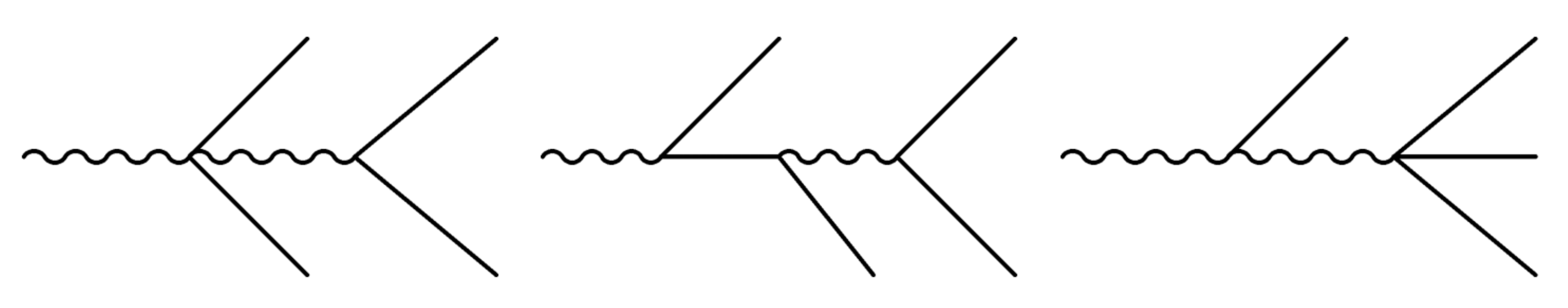} \caption{The first two Feynman diagrams to the $\pi^+\pi^+\pi^-\pi^-$ channel and the last Feynman diagram to the $\pi^0\pi^0\pi^+\pi^-$ channel} \label{Fig:EMC}
\end{figure}
For higher-order electromagnetic corrections, only tree diagrams are considered for simplicity. The relevant Feynman diagrams are shown in Fig.~\ref{Fig:EMC}.
For the third Feynman diagram in Fig.~\ref{Fig:EMC}, the leading order contribution is from the Wess-Zumino-Witten (WZW) anomaly \cite{Wess:1971yu,Witten:1983tw}. The Lagrangian at $\mathcal{O}(p^4)$ is given as \cite{Dai:2013joa, Wang:2023njt}, 
\begin{eqnarray}
\mathcal{L}^{GB}_{(4)}&=&\frac{\sqrt{2}N_C}{8\pi^2F}\epsilon_{\mu\nu\rho\sigma}\bigl<\Phi\partial^{\mu}v^{\nu}\partial^{\rho}v^{\sigma}\bigr> \nonumber\\
&+&i\frac{\sqrt{2}N_C}{12\pi^2F^3}\epsilon_{\mu\nu\rho\sigma}\bigl<\partial^\mu\Phi\partial^\nu\Phi\partial^\rho\Phi v^\sigma\bigr>,
\end{eqnarray}
with $v^\sigma$ the external vector current. The detailed results for the amplitudes of these three electromagnetic correction Feynman diagrams are given below. 
The form factors for $e^+e^-\rightarrow \pi^+\pi^+\pi^-\pi^-$ are: 
\begin{widetext}
\begin{equation}
J^\mu_{EMC,c}(p_1,p_2,p_3,p_4)=J^\mu_{EMC,1}+J^\mu_{EMC,2}\,,
\end{equation}
where the first part is given as 
\begin{align}
&J^\mu_{EMC,1}(p_1,p_2,p_3,p_4)=\!-\!2e^2\biggl\{\frac{2(p_1^\mu\!-\!p_3^\mu)}{\!-\!s_{12}\!+\!4m_\pi^2\!+\!2t_1\!+\!t_3\!-\!t_4\!+\!2\nu}\!+\!\frac{2(p_1^\mu\!-\!p_4^\mu)}{\!-\!s_{12}\!+\!4m_\pi^2\!+\!2t_1\!-\!t_3\!+\!t_4\!-\!2\nu}\notag\\
&\;\;\;\;\!+\!\frac{2(p_2^\mu\!-\!p_3^\mu)}{\!-\!s_{12}\!+\!4m_\pi^2\!+\!2t_2\!+\!t_3\!-\!t_4\!-\!2\nu}\!+\!\frac{2(p_2^\mu\!-\!p_4^\mu)}{\!-\!s_{12}\!+\!4m_\pi^2\!+\!2t_2\!-\!t_3\!+\!t_4\!+\!2\nu}\notag\\
&\;\;\;\;+\!\frac{p_1^\mu(\!-\!3s_{12}\!+\!4m_\pi^2\!+\!2t_1\!+\!4t_2\!-\!3t_3\!-\!t_4\!+\!2\nu)}{(\!-\!t_1\!+\!t_2\!+\!t_3\!+\!t_4)(\!-\!s_{12}\!+\!4m_\pi^2\!+\!2t_2\!+\!t_3\!-\!t_4\!-\!2\nu)}\!-\!\frac{p_1^\mu(\!3s_{12}\!-\!4m_\pi^2\!-\!2t_1\!-\!4t_2\!+\!t_3\!+\!3t_4\!+\!2\nu)}{(\!-\!t_1\!+\!t_2\!+\!t_3\!+\!t_4)(\!-\!s_{12}\!+\!4m_\pi^2\!+\!2t_2\!-\!t_3\!+\!t_4\!+\!2\nu)}\notag\\
&\;\;\;\;+\!\frac{p_2^\mu(\!-\!3s_{12}\!+\!4m_\pi^2\!+\!4t_1\!+\!2t_2\!-\!t_3\!-\!3t_4\!+\!2\nu)}{(\!t_1\!-\!t_2\!+\!t_3\!+\!t_4)(\!-\!s_{12}\!+\!4m_\pi^2\!+\!2t_1\!-\!t_3\!+\!t_4\!-\!2\nu)}\!-\!\frac{p_2^\mu(\!3s_{12}\!-\!4m_\pi^2\!-\!4t_1\!-\!2t_2\!+\!3t_3\!+\!t_4\!+\!2\nu)}{(\!t_1\!-\!t_2\!+\!t_3\!+\!t_4)(\!-\!s_{12}\!+\!4m_\pi^2\!+\!2t_1\!+\!t_3\!-\!t_4\!+\!2\nu)}\notag\\
&\;\;\;\;+\!\frac{p_3^\mu(\!3s_{12}\!-\!4m_\pi^2\!-\!2t_1\!+\!t_3\!-\!t_4\!+\!2\nu)}{(\!t_1\!+\!t_2\!-\!t_3\!+\!t_4)(\!-\!s_{12}\!+\!4m_\pi^2\!+\!2t_2\!-\!t_3\!+\!t_4\!+\!2\nu)}\!-\!\frac{p_3^\mu(\!-\!3s_{12}\!+\!4m_\pi^2\!+\!2t_2\!\!-\!t_3\!+\!t_4\!+\!2\nu)}{(\!t_1\!+\!t_2\!-\!t_3\!+\!t_4)(\!-\!s_{12}\!+\!4m_\pi^2\!+\!2t_1\!-\!t_3\!+\!t_4\!-\!2\nu)}\notag\\
&\;\;\;\;+\!\frac{p_4^\mu(\!3s_{12}\!-\!4m_\pi^2\!-\!2t_2\!-\!t_3\!+\!t_4\!+\!2\nu)}{(\!t_1\!+\!t_2\!+\!t_3\!-\!t_4)(\!-\!s_{12}\!+\!4m_\pi^2\!+\!2t_1\!+\!t_3\!-\!t_4\!+\!2\nu)}\!-\!\frac{p_4^\mu(\!-\!3s_{12}\!+\!4m_\pi^2\!+\!2t_1\!+\!t_3\!-\!t_4\!+\!2\nu)}{(\!t_1\!+\!t_2\!+\!t_3\!-\!t_4)(\!-\!s_{12}\!+\!4m_\pi^2\!+\!2t_2\!+\!t_3\!-\!t_4\!-\!2\nu)}\biggr\}\,,
\end{align}
and the second part is 
\begin{align}
 J^\mu_{EMC,2}(p_1,p_2,p_3,p_4)=&-\frac{4e^2}{F_\pi^2}\biggl\{\frac{(p_1^\mu-p_3^\mu)(8L_4^r(2m_K^2+m_\pi^2)+8L_5^rm_\pi^2+L_9^r(-s_{12}+4m_\pi^2+2t_1+t_3-t_4+2\nu))}{-s_{12}+4m_\pi^2+2t_1+t_3\!-\!t_4\!+\!2\nu}\notag\\
&+\frac{(p_1^\mu-p_4^\mu)(8L_4^r(2m_K^2+m_\pi^2)+8L_5^rm_\pi^2+L_9^r(-s_{12}+4m_\pi^2+2t_1-t_3+t_4-2\nu))}{-s_{12}+4m_\pi^2+2t_1-t_3+t_4-2\nu}\notag\\
&+\frac{(p_2^\mu-p_3^\mu)(8L_4^r(2m_K^2+m_\pi^2)+8L_5^rm_\pi^2+L_9^r(-s_{12}+4m_\pi^2+2t_2+t_3-t_4-2\nu))}{-s_{12}+4m_\pi^2+2t_2+t_3-t_4-2\nu}\notag\\
&+\frac{(p_2^\mu-p_4^\mu)(8L_4^r(2m_K^2+m_\pi^2)+8L_5^rm_\pi^2+L_9^r(-s_{12}+4m_\pi^2+2t_2-t_3+t_4+2\nu))}{-s_{12}+4m_\pi^2+2t_2-t_3+t_4+2\nu}\notag\\
&-\frac{1}{-s_{12}+4m_\pi^2+2t_2+t_3-t_4-2\nu}\bigl(-p_2^\mu(-8L_{10}^rt_3+8L_4^r(2m_K^2+m_\pi^2)\notag\\
&+8L_5^rm_\pi^2+L_9^r(-s_{12}+4m_\pi^2+2t_1+2t_2-3t_3+t_4-2\nu))\notag\\
&+p_3^\mu(-8L_{10}^rt_2+8L_4^r(2m_K^2+m_\pi^2)+8L_5^rm_\pi^2\notag\\
&+L_9^r(-s_{12}+4m_\pi^2+2t_1-2t_2+t_3+t_4-2\nu))+2L_9^r(t_2-t_3)p_1^\mu+2L_9^r(t_2-t_3)p_4^\mu\bigr)\notag\\
&+\frac{1}{-s_{12}+4m_\pi^2+2t_1+t_3-t_4+2\nu}\bigl(p_1^\mu(-8L_{10}^rt_3+8L_4^r(2m_K^2+m_\pi^2)\notag\\
&+8L_5^rm_\pi^2+L_9^r(-s_{12}+4m_\pi^2+2t_1+2t_2-3t_3+t_4+2\nu))\notag\\
&-p_3^\mu(-8L_{10}^rt_1+8L_4^r(2m_K^2+m_\pi^2)+8L_5^rm_\pi^2\notag\\
&+L_9^r(-s_{12}+4m_\pi^2-2t_1+2t_2+t_3+t_4+2\nu))-2L_9^r(t_1-t_3)p_2^\mu-2L_9^r(t_1-t_3)p_4^\mu\bigr)\notag\\
&+\frac{1}{-s_{12}+4m_\pi^2+2t_1-t_3+t_4-2\nu}\bigl(p_1^\mu(-8L_{10}^rt_4+8L_4^r(2m_K^2+m_\pi^2)+8L_5^rm_\pi^2\notag\\
&+L_9^r(-s_{12}+4m_\pi^2+2t_1+2t_2+t_3-3t_4-2\nu))\notag\\
&+p_4^\mu(8L_{10}^rt_1-8L_4^r(2m_K^2+m_\pi^2)-8L_5^rm_\pi^2\notag\\
&-L_9^r(-s_{12}+4m_\pi^2-2t_1+2t_2+t_3+t_4-2\nu))-2L_9^r(t_1-t_4)p_2^\mu-2L_9^r(t_1-t_4)p_3^\mu\bigr)\notag\\
&+\frac{1}{-s_{12}+4m_\pi^2+2t_2-t_3+t_4+2\nu}\bigl(p_2^\mu(-8L_{10}^rt_4+8L_4^r(2m_K^2+m_\pi^2)+8L_5^rm_\pi^2\notag\\
&+L_9^r(-s_{12}+4m_\pi^2+2t_1+2t_2+t_3-3t_4+2\nu))\notag\\
&+p_4^\mu(8L_{10}^rt_1-8L_4^r(2m_K^2+m_\pi^2)-8L_5^rm_\pi^2\notag\\
&-L_9^r(-s_{12}+4m_\pi^2+2t_1-2t_2+t_3+t_4+2\nu))-2L_9^r(t_2-t_4)p_1^\mu-2L_9^r(t_2-t_4)p_3^\mu\bigr)\notag\\
&+\frac{1}{(-t_1+t_2+t_3+t_4)(4m_\pi^2-s_{12}+2t_2+t_3-t_4-2\nu)}\bigl(\notag\\
&(8L_4^rm_K^2+4(l_4^r+L_5^r)m_\pi^2+L_9^rq^2)(4m_\pi^2-3s_{12}+2t_1+4t_2-3t_3-t_4+2\nu)p_1^\mu\notag\\
&+(8L_5^rm_\pi^2+8L_4^r(2m_K^2+m_\pi^2)+L_9^r(4m_\pi^2-s_{12}+2t_2+t_3-t_4-2\nu))\notag\\
&(4m_\pi^2-3s_{12}+2t_1+4t_2-3t_3-t_4+2\nu)p_1^\mu\bigr)\notag\\
&+\frac{1}{(-t_1+t_2+t_3+t_4)(4m_\pi^2-s_{12}+2t_2-t_3+t_4+2\nu)}\bigl(\notag\\
&(4m_\pi^2-3s_{12}+2t_1+4t_2-t_3-3t_4-2\nu)\notag\\
&(8L_5^rm_\pi^2+8L_4^r(2m_K^2+m_\pi^2)+L_9^r(4m_\pi^2-s_{12}+2t_2-t_3+t_4+2\nu))p_1^\mu\notag\\
&+(8L_4^rm_K^2+4(L_4^r+L_5^r)m_\pi+L_9^rq^2)(4m_\pi^2-3s_{12}+2t_1+4t_2-t_3-3t_4-2\nu)p_1^\mu\bigr)\notag\\
&+\frac{1}{(t_1-t_2+t_3+t_4)(4m_\pi^2-s_{12}+2t_1-t_3+t_4-2\nu)}\bigl(\notag\\
&(8L_4^rm_K^2+4(L_4^r+L_5^r)m_\pi^2+L_9^rq^2)(4m_\pi^2-3s_{12}+4t_1+2t_2-t_3-3t_4+2\nu)p_2^\mu\notag\\
&+(8L_5^rm_\pi^2+8L_4^r(2m_K^2+m_\pi^2))+L_9^r(4m_\pi^2-s_{12}+2t_1-t_3+t_4-2\nu)\notag\\
&(4m_\pi^2-3s_{12}+4t_1+2t_2-t_3-3t_4+2\nu)p_2^\mu\notag\bigr)\notag\\
&+\frac{1}{(t_1-t_2+t_3+t_4)(4m_\pi^2-s_{12}+2t_1+t_3-t_4+2\nu)}\bigl(\notag\\
&(4m_\pi^2-3s_{12}+4t_1+2t_2-3t_3-t_4-2\nu)\notag\\
&(8L_5^rm_\pi^2+8L_4^r(2m_K^2+m_\pi^2)+L_9^r(4m_\pi^2-s_{12}+2t_1+t_3-t_4+2\nu))p_2^\mu\notag\\
&+(8L_4^rm_K^2+4(L_4^r+L_5^r)m_\pi+L_9^rq^2)(4m_\pi^2-3s_{12}+4t_1+2t_2-3t_3-t_4-2\nu)p_2^\mu\bigr)\notag\\
&-\frac{1}{(t_1+t_2-t_3+t_4)(4m_\pi^2-s_{12}+2t_1-t_3+t_4-2\nu)}\bigl(\notag\\
&(8L_4^rm_K^2+4(L_4^r+L_5^r)m_\pi^2+L_9^rq^2)(4m_\pi^2-3s_{12}+2t_2-t_3+t_4+2\nu)p_3^\mu\notag\\
&+(8L_5^rm_\pi^2+8L_4^r(2m_K^2+m_\pi^2)+L_9^r(4m_\pi^2-s_{12}+2t_1-t_3+t_4-2\nu))\notag\\
&(4m_\pi^2-3s_{12}+2t_2-t_3+t_4+2\nu)p_3^\mu\bigr)\notag\\
&-\frac{1}{(t_1+t_2-t_3+t_4)(4m_\pi^2-s_{12}+2t_2-t_3+t_4+2\nu)}\bigl(\notag\\
&(4m_\pi^2-3s_{12}+2t_1-t_3+t_4-2\nu)\notag\\
&(8L_5^rm_\pi^2+8L_4^r(2m_K^2+m_\pi^2)+L_9^r(4m_\pi^2-s_{12}+2t_2-t_3+t_4+2\nu))p_3^\mu\notag\\
&+(8L_4^rm_K^2+4(L_4^r+L_5^r)m_\pi+L_9^rq^2)(4m_\pi^2-3s_{12}+2t_1-t_3+t_4-2\nu)p_3^\mu\bigr)\notag\\
&-\frac{1}{(t_1+t_2+t_3-t_4)(4m_\pi^2-s_{12}+2t_2+t_3-t_4-2\nu)}\bigl(\notag\\
&(8L_4^rm_K^2+4(L_4^r+L_5^r)m_\pi^2+L_9^rq^2)(4m_\pi^2-3s_{12}+2t_1+t_3-t_4+2\nu)p_4^\mu\notag\\
&+(8L_5^rm_\pi^2+8L_4^r(2m_K^2+m_\pi^2)+L_9^r(4m_\pi^2-s_{12}+2t_2+t_3-t_4-2\nu))\notag\\
&(4m_\pi^2-3s_{12}+2t_1+t_3-t_4+2\nu)p_4^\mu\bigr)\notag\\
&-\frac{1}{(t_1+t_2+t_3-t_4)(4m_\pi^2-s_{12}+2t_1+t_3-t_4+2\nu)}\bigl(\notag\\
&(4m_\pi^2-3s_{12}+2t_2+t_3-t_4-2\nu)\notag\\
&(8L_5^rm_\pi^2+8L_4^r(2m_K^2+m_\pi^2)+L_9^r(4m_\pi^2-s_{12}+2t_1+t_3-t_4+2\nu))p_4^\mu\notag\\
&+(8L_4^rm_K^2+4(L_4^r+L_5^r)m_\pi+L_9^rq^2)(4m_\pi^2-3s_{12}+2t_2+t_3-t_4-2\nu)p_4^\mu\bigr)\biggr\}\,.
\end{align} 
Notice that $J^\mu_{EMC,c}$ is the hadronization vector current for $\gamma^*\to\pi^+\pi^+\pi^-\pi^-$, and one does not need Eq.~(\ref{Eq:J;c}) to perform the transformation.
The only form factor of the hadronization vector current of EMC for $e^+e^-\rightarrow \pi^0\pi^0\pi^+\pi^-$ is from the third graph of Fig.~\ref{Fig:EMC}. One has 
\begin{equation}
\begin{aligned}
 J^\mu_{EMC,n}(p_1,p_2,p_3,p_4)&=\frac{e^2N_C^2}{576\pi^2F_\pi^4}\biggl\{\frac{1}{m_\pi^2+t_1-t_2+t_3+t_4}\bigl[((t_3-t_4)(s_{12}-2t_2+t_3+t_4)-2\nu(t_3+t_4))p_1^\mu\\
&+(-s_{12}(t_1+2t_4)+4m_\pi^2t_4+t_1(2t_2-t_3+t_4+2\nu))p_3^\mu\\
&+(s_{12}(t_1+2t_3)-4m_\pi^2t_3+t_1(-2t_2-t_3+t_4+2\nu))p_4^\mu\bigr]\\
&+\frac{1}{m_\pi^2-t_1+t_2+t_3+t_4}\bigl[((t_3-t_4)(s_{12}-2t_1+t_3+t_4)+2\nu(t_3+t_4))p_2^\mu\\
&-(s_{12}(t_2+2t_4)-4m_\pi^2t_4+t_2(-2t_1+t_3-t_4+2\nu))p_3^\mu\\
&+(s_{12}(t_2+2t_3)-4m_\pi^2t_3-t_2(2t_1+t_3-t_4+2\nu))p_4^\mu\bigr]\biggr\}.
\end{aligned}
\end{equation}
\end{widetext}
The pure cross section produced by the higher EMC corrections is about 0.006 percent compared with that of the ChPT ones shown in Fig.~\ref{Fig:cs}, which only contains the LO EMC contributions. The reason is the suppression of the electromagnetic coupling constant. One can conclude that the big discrepancy between the ChPT results and the cross section data can not be compensated by the higher-order electromagnetic corrections.

\section{The power-counting for the ChPT and RChT amplitudes}
As is known, it is hard to fix the power-counting for RChT. Nevertheless, the large $N_C$ expansion will compensate for it \cite{Pich:2002xy,Cirigliano:2006hb,Rosell:2004mn}. In this regard, the loop diagrams are suppressed by 1/$N_C$ compared with the tree diagrams, and in our Sol.~II, they can be ignored, since we only keep the LO $1/N_C$ terms \footnote{We are aware that in Refs.~\cite{Rosell:2004mn,Rosell:2006dt,Pich:2010sm,Nieves:2011gb}, the NLO $1/N_C$ corrections, including the ChPT loops, are taken into account.}.  
Indeed, the resummation of loops in meson-meson rescattering can dynamically generate resonances as intermediate states. A clear example is $\pi\pi$ scattering: when the ChPT amplitudes (including trees and loops) are unitarized as in Refs.~\cite{Dobado:1996ps,Dai:2011bs}, resonances like the $\sigma$ and $\rho$ appear as poles in the resulting amplitude. This implies a partial redundancy between the $\mathcal{O}(p^4)$ LEC terms and the resonance exchanges. 
Further, in our calculation, the momentum-dependent width of the $\rho$ are indeed derived from one-loop corrections of ChPT.
From this perspective, the loop contributions are partly double-counted with the resonance exchanges. 
Other than this, the integration of the heavier resonances will lead to higher derivative terms with pseudoscalar fields, which are double-counted with the higher order Lagrangians of ChPT except for the lowest order ones, e.g., $\mathcal{O}(p^2)$ terms. Therefore, the final remaining Feynman diagrams are the $\mathcal{O}(p^2)$ ChPT tree diagrams and the resonance tree diagrams in Sol.~II. 
For works following this method, see Refs.\cite{GomezDumm:2003ku,GomezDumm:2005nr,Dumm:2009kj,Dumm:2009va,Dai:2013joa}. 
For details about the large $N_C$ expansion and integrating out heavy resonances, see discussions below.

\subsection{Large $N_C$ in chiral perturbation theory}

Quantum corrections computed with the chiral Lagrangian will have a $1/N_C$ suppression for hadronic loop diagrams with pseudoscalars \cite{Pich:2002xy}. This can be observed clearly from the following expressions,
\begin{align}
F_\pi,F_V,G_V,c_d,c_m,\lambda_6^V,\lambda_{22}^V&\sim\mathcal{O}(\sqrt{N_C}), \nonumber\\
L_1,L_2,L_3,L_5,L_8,L_9,L_{10}&\sim\mathcal{O}(N_C),\nonumber\\
2L_1-L_2,L_4,L_6,\lambda_6^{VV},\lambda_3^{SV}&\sim\mathcal{O}(1),\nonumber\\
L_7&\sim\mathcal{O}(N_C^2),\nonumber\\
\lambda^{SVV}&\sim\mathcal{O}(1/\sqrt{N_C}).
\end{align}
Since ChPT one-loop amplitudes have a factor $1/F_\pi^4$, they are suppressed by $1/N_C^2$, while the tree-level amplitudes, either from $\mathcal{O}(p^2)$ ChPT Lagrangians, or from the RChT Lagrangians, are suppressed by factors of only $1/N_C$.Hence, we do not include the ChPT one-loop diagrams. 
For the tree-level diagrams generated by the $\mathcal{O}(p^4)$ ChPT Lagrangians, the $L_1,L_2,L_3,L_5,L_8,L_9,L_{10}$ amplitudes will be suppressed by $1/N_C$, too. However, these diagrams are double-counted with those tree-level resonance diagrams once the resonances are integrated out. See discussions below.

\subsection{Integrating out heavy fields at tree level}
\begin{figure}[!htb]
\centering
\includegraphics[width=0.48\textwidth]{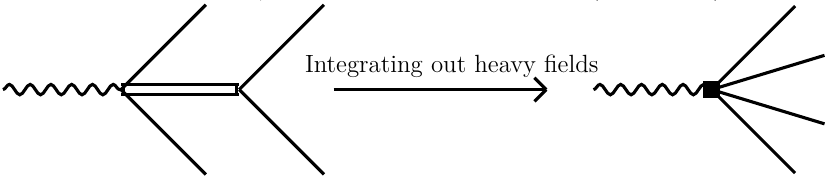} 
\caption{The example of integrating out heavy fields.}
\label{Fig:reply3}
\end{figure}

Integrating out a heavy field $H$ is through the following expressions,\\
\begin{equation}
 Z[l_i]=e^{iW_{\rm eff}[l_i]}\equiv\int [dH]e^{i\int d^4x\mathcal{L}(H(x),l_i(x))},
\end{equation}
leaving only the light fields $l_i$. 
In our analysis, the complete Lagrangian is given as 
\begin{equation} \mathcal{L}=\mathcal{L}^{\rm ChPT}+\mathcal{L}^{\rm RChT}=\mathcal{L}^{\rm ChPT}+\mathcal{L}_{kin}^{R}+\mathcal{L}_{int}^R.
\end{equation}
Here, $\mathcal{L}^{\rm ChPT}$ includes only the lowest order ChPT Lagrangians. 
As an example, we perform the resonance Lagrangians without scalars. 
The relevant full Lagrangians are,
\begin{eqnarray} 
\mathcal{L}_{kin}^{S}+\mathcal{L}_{int}^S&=&\!\bigl<-\frac{1}{2}S\mathcal{D}_SS+SJ_S\bigr>\nonumber\\
&=&\!\frac{1}{2}\left\langle \nabla^{\mu}S\nabla_{\mu}S\right\rangle\!-\!\frac{1}{2}M_{S}^{2}\left\langle SS\right\rangle
\!+\!c_d\bigl<Su_\mu u^\mu\bigr>\nonumber\\
&\!+\!&c_m\bigl<S\chi_+\bigr>
\!+\!\lambda^S_{15}\bigl<Sf_{+\mu\nu}f_+^{\mu\nu}\bigr>\nonumber\\
&\!+\!&\lambda_3^{SV}\bigl<\{S,V_{\mu\nu}\}f_+^{\mu\nu}\bigr>
+\lambda^{SVV}\bigl<SV_{\mu\nu}V^{\mu\nu}\bigr> \,, \nonumber\\
\end{eqnarray}
where one has 
\begin{eqnarray}
\mathcal{D}_S(x)&=&\nabla_{\mu}\nabla^{\mu}+M_S^2,\nonumber\\
 J_S(x)&=&c_du_\mu(x) u^\mu(x)+c_m\chi_+(x)+\lambda^S_{15}f_{+\mu\nu}(x)f_+^{\mu\nu}(x)\nonumber\\
&+&\lambda_3^{SV}[V_{\mu\nu}(x)f_+^{\mu\nu}(x)+f_+^{\mu\nu}(x)V_{\mu\nu}(x)]\nonumber\\
&+&\lambda^{SVV}V_{\mu\nu}(x)V^{\mu\nu}(x),\nonumber\\
\nabla^{\mu}A&=&\partial_\mu+[\Gamma_\mu,A],\nonumber\\
\Gamma_\mu&=&\{u^\dagger(\partial_\mu-ir_\mu)u+u(\partial_\mu-il_\mu)u^\dagger\}/2. \nonumber
\end{eqnarray}
Then, the interaction action can be written as
\begin{eqnarray}
&&\int d^4x\mathcal{L}_S(S,J_S)\nonumber\\
&=&\!\int\! d^4x\bigl<-\frac{1}{2}S\mathcal{D}_SS+SJ_S\bigr>\nonumber\\
&=&\!-\frac{1}{2}\!\int\! d^4x\bigl<(S\!-\!\mathcal{D}_S^{-1}J_S)\mathcal{D}_S(S\!-\!\mathcal{D}_S^{-1}J_S)\!-\!J_S\mathcal{D}_S^{-1}J_S\bigr>\nonumber\\
&=&\!-\frac{1}{2}\!\int\! d^4x\bigl<S'\mathcal{D}_SS'-J_S\mathcal{D}_S^{-1}J_S\bigr>
\end{eqnarray}
with
\begin{eqnarray}
& S'(x)=S(x)+\int d^4y\Delta_F(x-y)J_S(y),\nonumber\\
& (\nabla_{\mu}\nabla^\mu+M_S^2)\Delta_F(x-y)=-\delta^4(x-y). \nonumber
\end{eqnarray}
By integrating out the scalars, one has 
\begin{eqnarray}
 Z[J_S]&=&e^{iW_{\rm eff}[J_S]}\equiv\int [dS']e^{i\int d^4x\mathcal{L}(S'(x),J_S(x))}\nonumber\\
&=&Z[0]e^{\frac{i}{2}\int d^4x\bigl<J_S\mathcal{D}_S^{-1}J_S\bigr>},
\end{eqnarray}
where $Z[0]$ is an overall constant that does not have scalars,
\begin{equation}
 Z[0]=e^{i\int d^4x[\mathcal{L}^{\rm ChPT}+\mathcal{L}_{kin}^{S\!\!\!/}+\mathcal{L}_{int}^{S\!\!\!/}]}\int[dS']e^{i\int d^4x\bigl<-\frac{1}{2}S'\mathcal{D}_SS'\bigr>}. \nonumber
\end{equation}
Here, $S\!\!\!/$ means the resonance Lagrangians without scalars. 
$Z[0]$ can be dropped from further consideration. One can define the following effective action, 
\begin{equation}
 W_{\rm eff}[J_S]=-\frac{1}{2}\int d^4xd^4yJ_S(x)\Delta_F(x-y)J_S(y). \nonumber
\end{equation}
The heavy scalar propagator exhibits exponentially localized behavior at short-distance scales (on the order of the $1/M_H$). This property enables the Taylor expansion of the non-local current operator $J_S(y)$ around a fixed point x:
\begin{equation}
 J_S(y)=J_S(x)+(y-x)^\mu[\partial_\mu J_S(y)]_{y=x}+\mathcal{O}\bigl((x-y)^2\bigr). \nonumber
\end{equation}
thereby transforming the original non-local action into a series expansion of local operators.
Keeping the leading term and applying the following equation, 
\begin{equation}
 \mathcal{D}^{-1}_S(x)J(x)=-\int d^4y\Delta_F(x-y)J(y).
\end{equation}
One can obtain 
\begin{equation}
 \int d^4y \Delta_F(x-y)=-\frac{1}{M_S^2}, 
\end{equation}
and
\begin{eqnarray}
 W_{eff}[J_S]\!&=&\!\int\! d^4x\frac{1}{2M_S^2}\bigl<J_S(x)J_S(x)\bigr>\!+\mathcal{O}(M_S^{-4})\nonumber\\
\!&=&\!\int\! d^4x\frac{1}{2M_S^2}\bigl<\bigl(c_du_\mu(x) u^\mu(x)\!+\!c_m\chi_+(x)\bigr)^2\bigr>\nonumber\\
&&+\mathcal{O}(p^6)+\mathcal{O}(M_S^{-4})
\end{eqnarray}
Obviously, the leading interacting Lagrangians are $\mathcal{O}(p^4)$. This can be easily indicated by several Feynman diagrams, as shown in Fig.~\ref{Fig:reply3}. 
One can integrate out the scalar meson at tree level in the left diagram, 
and it will generate out an effective vertex, $\bigl<[c_du_\mu(x) u^\mu(x)+c_m\chi_+(x)]^2\bigr>$, which is $\mathcal{O}(p^4)$ in the chiral counting. See the right side diagram in Fig.~\ref{Fig:reply3}. 
These effective Lagrangians may be double-counted with the $\mathcal{O}(p^4)$ ChPT Lagrangians with the LECs $L_i$.
For other resonances, one has similar conclusions.
This is why the $\mathcal{O}(p^4)$ ChPT Lagrangians are omitted when applying RChT calculation.
Indeed, our strategy is the same as what is done in Refs.~\cite{Ecker:1988te,GomezDumm:2003ku,GomezDumm:2005nr,Dumm:2009kj,Dumm:2009va}, where the ChPT Lagrangian is only used up to $O(p^2)$, and the resonances are included explicitly.

It should be noticed that in the discussions above, the two operations, integrating out resonances and trace, can be exchanged. 
It can be checked by two performances: one performs the trace of the scalar octet, $\int d^4x\bigl<-\frac{1}{2}S\mathcal{D}_SS+SJ_S\bigr>$ at first, and next the integrating out scalars; or vice versus, integrating out the scalars, then the trace, $\int d^4x\frac{1}{2m_S^2}\bigl<J_S(x)J_S(x)\bigr>$. 
As can be found, both methods give the same results. For integrating out the $\sigma$ field in both ways, one has the same effective Lagrangians, 
\begin{eqnarray}
\mathcal{L}^{\rm eff} &=&\frac{1}{2M_\sigma^2}\biggl[-\frac{2\sqrt{2}c_mm_\pi^2\pi^-\pi^+}{F^2}
 +\frac{2\sqrt{2}c_d}{F^2}\bigl(\partial_\mu\pi^-\partial^\mu\pi^+ \nonumber\\
&&+ie\pi^+A_\mu\partial^\mu\pi^--ie\pi^-A_\mu\partial^\mu\pi^+\bigr)\biggr]^2+\cdot\cdot\cdot
\end{eqnarray}
For other resonances, it will be the same.

\bibliography{reference}

\end{document}